\documentclass[draft]{agujournal2019}
\usepackage{url}
\usepackage{lineno}

\makeatletter
\providecommand\@jnl@name[1]{\def\jnl@name{#1}\jnl@name}
\newcommand\aj{\@jnl@name{AJ}}                  
\newcommand\araa{\@jnl@name{ARA\&A}}             
\newcommand\apj{\@jnl@name{ApJ}}               
\newcommand\apjl{\@jnl@name{ApJ}}             
\newcommand\apjs{\@jnl@name{ApJS}}            
\newcommand\ao{\@jnl@name{Appl.~Opt.}}        
\newcommand\apss{\@jnl@name{Ap\&SS}}          
\newcommand\aap{\@jnl@name{A\&A}}             
\newcommand\aapr{\@jnl@name{A\&A~Rev.}}       
\newcommand\aaps{\@jnl@name{A\&AS}}           
\newcommand\azh{\@jnl@name{AZh}}              
\newcommand\baas{\@jnl@name{BAAS}}            
\newcommand\icarus{\@jnl@name{Icarus}}        
\newcommand\jrasc{\@jnl@name{JRASC}}          
\newcommand\memras{\@jnl@name{MmRAS}}         
\newcommand\mnras{\@jnl@name{MNRAS}}          
\newcommand\pra{\@jnl@name{Phys.~Rev.~A}}     
\newcommand\prb{\@jnl@name{Phys.~Rev.~B}}
\newcommand\prc{\@jnl@name{Phys.~Rev.~C}}
\newcommand\prd{\@jnl@name{Phys.~Rev.~D}}
\newcommand\pre{\@jnl@name{Phys.~Rev.~E}}
\newcommand\prl{\@jnl@name{Phys.~Rev.~Lett.}} 
\newcommand\pasp{\@jnl@name{PASP}}            
\newcommand\pasj{\@jnl@name{PASJ}}            
\newcommand\qjras{\@jnl@name{QJRAS}}          
\newcommand\skytel{\@jnl@name{S\&T}}          
\newcommand\solphys{\@jnl@name{Sol.~Phys.}}   
\newcommand\sovast{\@jnl@name{Soviet~Ast.}}   
\newcommand\ssr{\@jnl@name{Space~Sci.~Rev.}}  
\newcommand\zap{\@jnl@name{ZAp}}              
\newcommand\nat{\@jnl@name{Nature}}           
\newcommand\iaucirc{\@jnl@name{IAU~Circ.}}    
\newcommand\aplett{\@jnl@name{Astrophys.~Lett.}}
\newcommand\apspr{\@jnl@name{Astrophys.~Space~Phys.~Res.}}
\newcommand\bain{\@jnl@name{Bull.~Astron.~Inst.~Netherlands}}
\newcommand\fcp{\@jnl@name{Fund.~Cosmic~Phys.}}
\newcommand\gca{\@jnl@name{Geochim.~Cosmochim.~Acta}}
\newcommand\grl{\@jnl@name{Geophys.~Res.~Lett.}}   
\newcommand\jcp{\@jnl@name{J.~Chem.~Phys.}}
\newcommand\jgr{\@jnl@name{J.~Geophys.~Res.}}       
\newcommand\jqsrt{\@jnl@name{J.~Quant.~Spec.~Radiat.~Transf.}}
\newcommand\memsai{\@jnl@name{Mem.~Soc.~Astron.~Italiana}}
\newcommand\nphysa{\@jnl@name{Nucl.~Phys.~A}}
\newcommand\physrep{\@jnl@name{Phys.~Rep.}}
\newcommand\physscr{\@jnl@name{Phys.~Scr}}
\newcommand\planss{\@jnl@name{Planet.~Space~Sci.}}
\newcommand\procspie{\@jnl@name{Proc.~SPIE}}
\newcommand\actaa{\@jnl@name{Acta~Astron.}}
\newcommand\caa{\@jnl@name{Chinese~Astron.~Astrophys.}}
\newcommand\cjaa{\@jnl@name{Chinese~J.~Astron.~Astrophys.}}
\newcommand\jcap{\@jnl@name{J.~Cosmology~Astropart.~Phys.}}
\newcommand\na{\@jnl@name{New~A}}
\newcommand\nar{\@jnl@name{New~A~Rev.}}
\newcommand\pasa{\@jnl@name{PASA}}
\newcommand\rmxaa{\@jnl@name{Rev.~Mexicana~Astron.~Astrofis.}}
\newcommand\maps{\@jnl@name{Meteorit.~Planet.~Sci.}}
\newcommand\aca{\@jnl@name{Astrochim.~Acta}}
\makeatother

\usepackage{xcolor}

\journalname{Space Weather}

\begin{document}

\title{Geometric tracking of a halo CME using full-circle wide-field views from PUNCH}

\authors{Craig~E. DeForest\affil{1},
Kieran~A. Russell\affil{2},
Samuel Van~Kooten\affil{1},
Heather~A. Elliott\affil{3},
Maher~A. Dayeh\affil{3},
Sarah~E. Gibson\affil{4},
Anna Malanushenko\affil{4},
David~F. Webb\affil{5},
Ronnie Killough\affil{3},
Jillian Redfern\affil{1},
William~L. Wells\affil{3},
Matthew~N. Beasley\affil{1},
Glenn~T. Laurent\affil{1},
Nick~R. Waltham\affil{6},
J.~Marcus Hughes\affil{1},
Daniel~B. Seaton\affil{1},
Ritesh Patel\affil{1},
Chris Lowder\affil{1},
Derek~A. Lamb\affil{1},
Alec~R. DeForest\affil{7}
}

\affiliation{1}{Southwest Research Institute, Boulder CO, USA}
\affiliation{2}{Michigan State University, East Lansing MI, USA}
\affiliation{3}{Southwest Research Institute, San Antonio TX, USA}
\affiliation{4}{National Center for Atmospheric Research / High Altitude Observatory, Boulder CO, USA}
\affiliation{5}{ISR/Boston College, Chestnut Hill, MA, USA}
\affiliation{6}{UK Research \& Innovation / Science \& Technology Facilities Council / Rutherford Appleton Laboratory, Oxfordshire, UK}
\affiliation{7}{Independent researcher, Nederland CO, USA}

\correspondingauthor{Craig~E. DeForest}{craig.deforest@swri.org}

\begin{keypoints}
\item PUNCH tracked a fast halo CME continuously from the Sun to Earth impact using visible light imaging out to solar elongation of 47$^\circ$. 
\item An extremely simple three parameter geometry model captured the CME off-axis geometry and its increasing angular speed with distance.
\item Forecasts using the simple model converged identifiably to high accuracy of 30$\pm$ 15 minutes of the CME impact, 8 hours before arrival.
\end{keypoints}

\begin{abstract}

We tracked the large, geoeffective halo coronal mass ejection (CME) of 2025-May-31 across the 
very wide field of view of the Polarimeter to Unify the Corona and Heliosphere (PUNCH) mission,
as a proof of concept for data-informed forecasting of CME arrival times at Earth. 
Using simple point-and-click measurement of the front location as far as 47$^\circ$ away from the Sun itself, we constrained an analytic "ludicrous ice-cream cone" geometric model with just three parameters: position angle, off-Earth-Sun angle, and propagation speed. We show that the model is sufficient to capture the bulk envelope of the CME, and also insufficient to capture the complexity of the CME's evolution. We generated simulated arrival-time forecasts for the CME by extrapolating the constant-velocity propagation to
determine the time of impact with Earth.  These ``retroforecasts'' converged
to within $30\pm15$ minutes of the actual arrival time, 8 hours prior to arrival.
The convergence was marked by a clear inflection point in the curve of predicted-arrival-time vs. time-of-forecast.  We conclude that continuous wide-field tracking is sufficiently powerful to enable even crude models to substantally improve arrival-time forecasting precision, and discuss future work required both to validate this single-event result and to extend it to more sophisticated analyses.

\end{abstract}

\section*{Plain Language Summary}

The Sun regularly launches vast clouds of magnetized plasma called coronal mass
ejections (CMEs), which can disrupt satellites, power grids, and communications
when they strike Earth.  Forecasting exactly when a CME will arrive has been difficult
because existing instruments lose sight of the CME long before it reaches us.
The PUNCH mission images a wide swath of the sky and, in late May 2025,
was able to follow the mission's first fast Earth-directed CME almost the entire way from the 
Sun to Earth.  By
tracing the front of the cloud by hand in the images and fitting a
simple geometric model, we produced arrival-time ``retroforecasts'' that improved as the
CME approached and ultimately settled within $30\pm 15$ minutes of the true impact 
time, much more precise than contemporary forecasts for that event. We identified 
an indicator, 
available from the forecasts 
themselves, that shows when the forecast has converged to a stable value.  The overall result 
depends only weakly on built-in assumptions in the model itself. 
The result demonstrates explicitly that continuous wide-field imaging from near Earth can substantially improve 
space weather forecasts.

\section{Introduction} 

Fast coronal mass ejections (CMEs) are the main source of strong space weather events at Earth 
\cite<e.g.,>{gosling_1993,Baker_1998}, and forecasting the arrival time and impact aspect of CMEs (i.e., 
direct, off-center, or flank) is a
major driver of research in space weather physics \cite<e.g.,>{webb_howard_2012,moldwin_2022book}. CME fronts, like 
other dense structures in the corona and solar wind, may be imaged directly via Thomson scattering
of sunlight from free electrons contained in the structure. They have been studied via image 
analysis since the advent of externally occulted coronagraphs such as those on OSO-7 \cite{koomen_etal_1975}
and Skylab \cite{macqueen_etal_1974}. Earth-directed CMEs appear as halos in coronagraph images 
\cite{howard_etal_1982}. Since the launch of SOHO/LASCO 
\cite{brueckner_etal_1995}, the corona has been routinely monitored to elongation angles up to 7.5$^\circ$ from the Sun itself, in part to identify and forecast the arrival of geoeffective CMEs 
\cite<e.g.,>{schwenn_06}.  

Coronagraph images from near Earth are limited for forecasting CME arrival time and 
geoeffectiveness, because of two major challenges
that are addressed by wider-field heliospheric imaging: (1) in the coronagraph field of view, 
there is a strong
ambiguity between CME angular width (relative to the Sun) and absolute speed, limiting 
the effectiveness of image sequences to characterize the CME's trajectory; 
and (2) viewer-directed 
CMEs leave the coronagraph field of view early in their trajectory,
requiring extrapolation across roughly 80\% of their trajectory before impact with Earth.
The first effect has 
been extensively studied in the literature.  Notable recent examples include 
\citeA{verbeke_etal2023}, who investigated the effect of the ambiguity in the coronagraph
field of view, both from a single viewpoint and multiple viewpoints, by studying 
manual fits to a simulated CME with known geometry; and \citeA{nikou_etal2025}, who 
investigated the effect on multiple coronagraphic viewpoints on forecast skill,
using real observational data. The second was highlighted by \citeA{kay_etal2020}, who found
that both small errors in drag modeling and small errors in CME width estimation can lead 
to significant (many hour) uncertainty in CME arrival time forecasts from coronagraphic measurements.

Fortunately, CMEs are not only visible in the solar corona. They may be tracked 
with visible light across the solar system, even to distances more than 1 AU.  Wide 
field imaging both breaks the ambiguity and permits longer tracking of the CME as it
propagates \cite<e.g.,>{deforest_etal_2016}. From a forecast perspective,
such wide-field imagers offer twin advantages: firstly, they afford tracking 
CMEs directly over a larger fraction of their propagation from Sun to Earth
\cite{deforest_etal_2016}; and secondly, they allow tracking strong CMEs 
outside of the particle storms that are frequently observed near the launch of the
strongest CMEs, potentially adding resilience to the data stream used for space weather forecasting.

The first wide-field visible measurements of CMEs used the three photometers on the 
spinning Helios spacecraft to reconstruct crude wide-field images of CMEs as they passed through
the inner solar system \cite{jackson_leinert_1985}. Later, both the SMEI \cite{eyles_etal_2003} and
STEREO/SECCHI \cite{howard_etal_2008} instruments were used to further develop wide-field (heliospheric)
imaging of CMEs and the solar wind.  SMEI offered essentially full-sphere imaging of the sky, 
and was used extensively to track CMEs despite limited signal-to-noise ratio; key references
include \citeA{tappin_etal_2004, webb2006solar, howard2006tracking}.  SMEI in particular 
demonstrated the importance of combining visible imaging data with 
interplanetary radio scintillation measurements, to exploit the advantages of both
\cite{jackson2007comparison}.
Meanwhile, STEREO imaging from off the Earth-Sun line has been used to study CME propagation 
\cite<e.g.,>{webb_etal2009a} and, in combination with in-situ sampling of the CME plasma, CME 
evolution and origin \cite{deforest_etal_2013a}; and wide-field measurements from near the Earth-Sun line, augmented by off-axis measurements, have been shown to be useful in forecasting CMEs \cite{amerstorfer_etal2018}.  
More recent instruments specifically designed to record Thomson-scattered light at high solar 
elongation angles include SoloHI \cite{howard_etal_2020} and WISPR \cite{vourlidas_etal_2016}, 
and recent work has shown that off-axis triangulation from SoloHI can improve estimation of CME
location in 3-D for distances as large as 75 $r_\odot$ (0.35 AU) from the Sun \cite{paouris_etal2025}.

The PUNCH mission \cite{deforest2026} comprises one coronagraphic Narrow Field Imager \cite<NFI; >{colaninno_2025} and 
three heliospheric-imager Wide Field Imagers \cite<WFI; >{laurent_2026}, working together as a single ``virtual instrument''.
PUNCH was specifically designed to image CMEs and other solar 
wind features in more detail and over more position angles than could STEREO, and with more 
sensitivity and spatiotemporal resolution than could SMEI.
PUNCH maintains a circular, 90$^\circ$ wide field of view centered on 
the Sun, to track CMEs as they cross the inner solar system well beyond the LASCO field of view, 
which 
has a diameter of approximately $15^\circ$ centered on the Sun. 
Understanding CME propagation is an explicit
science objective of the mission. PUNCH is intended to locate features in 3-D using the
polarimetric properties 
of Thomson scattering, but even before polarimetric data were available, its large 
field of view and high resolution addressed both the narrow-field ambiguity and extrapolation challenges
faced by coronagraphs, potentially enabling 3-D reconstruction and improved arrival time forecasting
using wide-angle geometric effects.
In this paper, we investigate the feasibility of using the improved wide-angle imaging with PUNCH or similar
instrumentation, to improve
arrival-time forecasts of fast halo CMEs.

A few days after the PUNCH data acquisition campaign began in 2025 May, a 
large halo
CME erupted just before the start of 2025 May 31. 
It was bright enough to be seen even with preliminary data processing on uncalibrated PUNCH data.  
The CME crossed from Sun to Earth in 30 hours and proved to be geoeffective.  We chose this 
early CME to study the effect of wide-field tracking 
on arrival-time forecasting. This was a nice observation for such a test: the preliminary data set 
shows the evolving front but is severely limited in quality compared to later iterations of the
PUNCH data, isolating the effect of rapid (8-30 minute cadence), moderate resolution (few-arcmin) 
wide-field imaging from later improvements in signal-to-noise ratio and/or availability of polarimetry.

We tracked the CME front and fitted the resulting front location data to a simplified geometric model inspired
by earlier analytic models \cite{xue_etal2005,tappin_howard_2009,na2017development}, and 
used analytic propagation to estimate arrival time of the CME at Earth, refining this mock forecast
(``retroforecast'') every 30 minutes.  In this respect we followed in the footsteps of
\citeA{webb_etal2009b}, 
who used SMEI front-geometry data, to produce refined forecasts every 90-180 minutes of simulated
time. We found that the overall propagation, coupled with frequent updates, was sufficient to 
generate a high quality arrival time forecast that converged to within $30\pm15$ minutes of the
actual arrival time, some 8 hours before impact.  Importantly: the convergence was directly identifiable in the 
sequence of mock forecast times, with a sharp inflection point that marked a transition to 
stable forecasts as new data arrived. 

In the remainder of this article, we describe the data themselves (Section \ref{sec:data}), 
outline the method we used to analyze the CME propagation, including the very simplified
three-free-parameter geometric model (Section 
\ref{sec:methods}), present the results of applying a forecasting method based on 
that model (Section \ref{sec:results}), and discuss and draw conclusions (including future follow-on work needed to validate, develop, and apply the 
result; Section \ref{sec:conclusion}).

\begin{figure}
    \centering
    \includegraphics[width=0.9\linewidth]{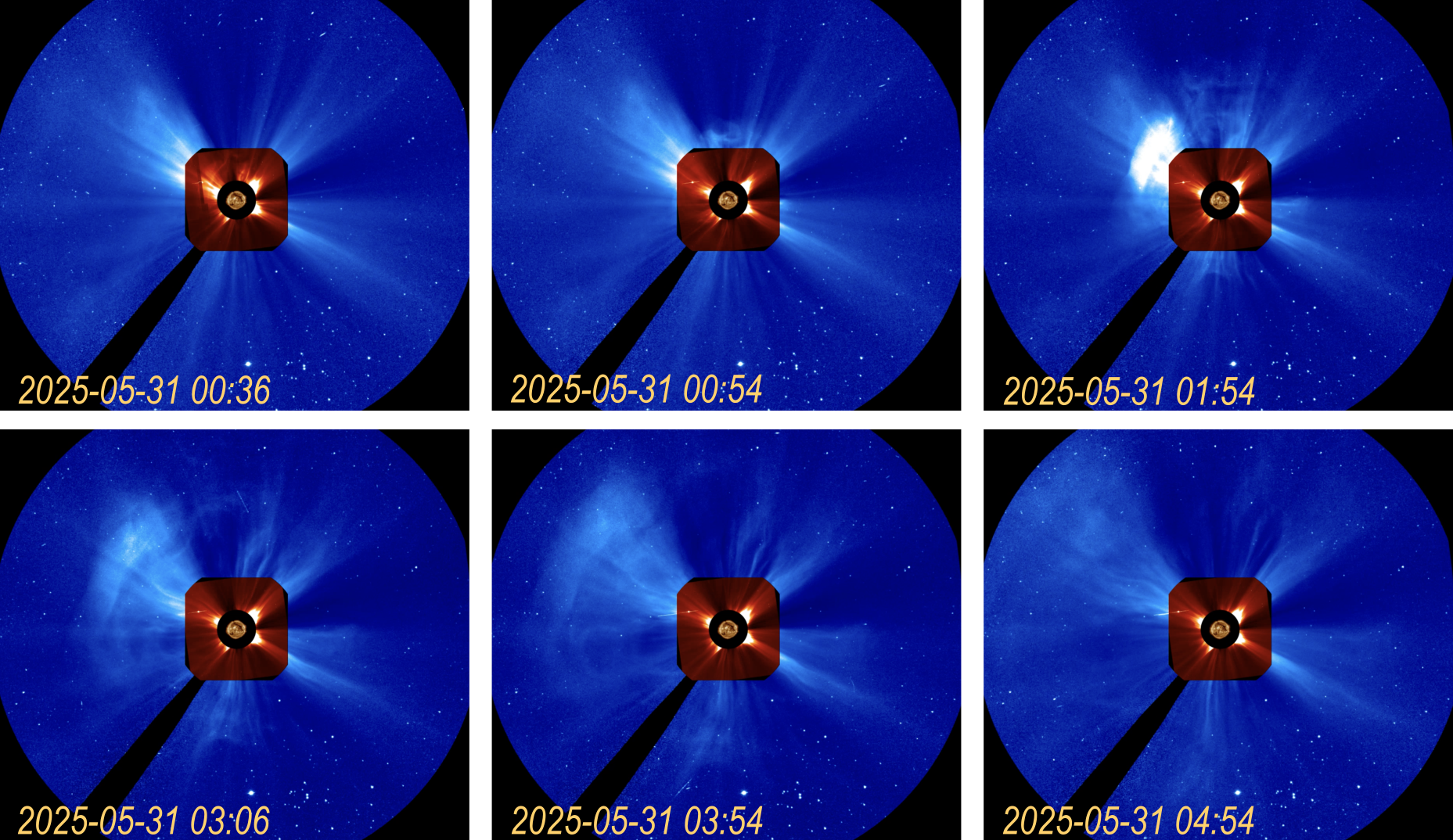}
    \caption{Composite images from SOHO/LASCO C2 and C3 show the early progress of the 2025-May-31 CME, associated with the flare of 2025 May 30 23:30.  The first panel shows less than 1/2 hour of progress; subsequent panels are at approximately 1-hour intervals.  The CME crossed the C2 field of view in under one hour, and the C3 field of view in five hours.  (Images: Helioviewer Project)}
    \label{fig:lasco}
\end{figure}

\begin{figure}
    \centering
    \includegraphics[width=0.75\linewidth]{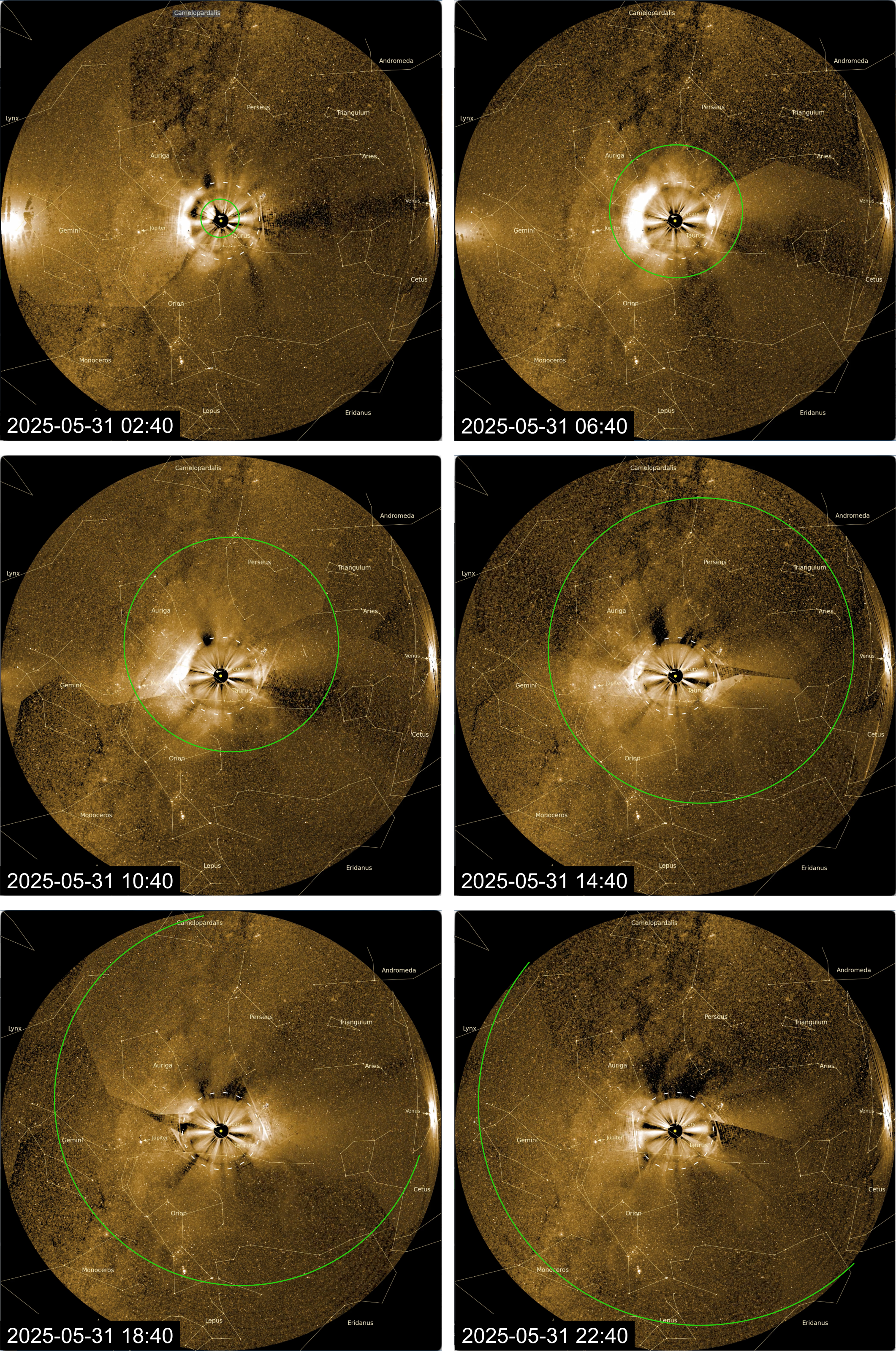}
    \caption{Six full-field PUNCH mosaics from the sample data and after preliminary processing, described in Section \ref{sec:processing}, demonstrate the appearance of the halo CME in the early PUNCH data. The field of view is over 90$^\circ$ on the sky. For scale, the constellation Orion is visible at lower left; and the NFI field of view
    (matching the LASCO C-3 field rendered in Figure \ref{fig:lasco}) is marked with a dashed circle. Because the boundary can be difficult to pick out in the individual frames, the rough location of the CME front is marked with a green arc.  The actual leading edge is inside the green arc in all cases.  The front may be easier to see in the accompanying movie (in the supporting information to this article).}
    \label{fig:punch-mosaics}
\end{figure}

\section{Data} \label{sec:data}

On 2025 May 31, a halo
CME left the Sun and propagated toward Earth. Preliminary processing reduced the stellar background 
and F corona sufficiently to reveal morphology of the evolving CME front. 
Figure \ref{fig:punch-mosaics} shows six frames from this data set. In the following subsections,
we describe the data themselves (\ref{sec:processing}), and the halo CME that we observed and
analyzed (\ref{sec:halo-cme}).

\subsection{PUNCH Initial Release images}\label{sec:processing}

We analyzed early-release data from PUNCH, acquired shortly after the regular observing 
sequence began and well before the spacecraft had reached their nominal orbital configuration of 
120$^\circ$ separation in orbital anomaly. The only stage of the data pipeline that was fully commissioned
was 
Level 0 (``L0''; raw camera images assembled into data files), with very preliminary Level 1
(``L1''; photometric) processing based on calibration that was loosely characterized pre-launch
and intended to be refined in-flight using celestial sources. 
We advanced these preliminary L1 data to Level 3 (``L3''; background-subtracted mosaics) using prototype versions of the
ultimate pipeline code
described by \citeA{deforest2026} and references therein.  We created preliminary L2 and L3 images 
at 2$\times$2 binned resolution, for an overall size of 2k$\times$2k pixels The binning used $3\times 3$ median filtering on the original mosaics, to remove the bright cores of stars from the averages.  Further, The 
Narrow Field Imager (NFI) camera
suffered from 
bright variable terrestrial stray light, which we minimized using a principal-component
analysis method 
applied to the time series of NFI clear images.  Polarization data were not yet available,
so we used the mission's eight-minute cadence clear (unpolarized) data instead.

The preliminary L3 images used simple running-minimum subtraction in PUNCH's azimuthal equidistant helioprojective 
coordinate system, to 
reduce the effects of the F corona.  Further, the starfield was detected and partially suppressed using 
local-neighborhood
median filtering: each pixel was compared to the local median of its $5\times 5$ pixel neighborhood in the 
$2k\times 2k$ mosaic image plane.  Pixels that
exceeded their neighborhood median value by a factor of 3 were replaced by the median. 
This suppressed both bright stars and detector streak artifacts from saturation of the 
very brightest objects. The two median steps together reduced the brightness of the 
background starfield by over an order of magnitude, though artifacts remained due to the incomplete-at-that-time cross-instrument photometric calibration.

After preliminary-pipeline processing, the 8-minute-cadence images were converted to PNG display format in the PUNCH (gold) 
standard
color table. We raised the linear radiance pixels from PUNCH to the 0.25 power, for an overall data-display gamma 
of $(2.2)(0.25)=0.55$, when using a typical sRGB display \cite{anderson1996srgb} with gamma of 2.2.  (Recall that the luminance of most
digital displays scales as $L = k P^\gamma$, where: $L$ is the luminance of a particular pixel; $k$ is a calibration constant for the 
particular display; $P$ is the encoded value of that particular pixel; and $\gamma$ is the gamma value of the display). This fourth-root 
operation helped highlight faint regions at the cost of slightly reducing contrast.  These steps yielded a series of directly-viewable 
image frames
at the PUNCH clear-frame cadence of 8 minutes, forming a movie.  A sample of six of these preliminary frames, incorporating the 
complete PUNCH constellation and using a persistence-of-vision algorithm to
fill in the full field of view, is in Figure \ref{fig:punch-mosaics}; and the full movie is available in the supplementary 
information to this article and in the analysis' Zenodo repository \cite{deforest_etal_2026c}. 
The ``persistence-of-vision'' reduces distracting artifacts associated with the rotating trefoil shape of the PUNCH field of 
view, by propagating prior frames' pixel values into subsequent frames in which those values are missing.  

\begin{figure}
    \centering
    \includegraphics[width=0.75\linewidth]{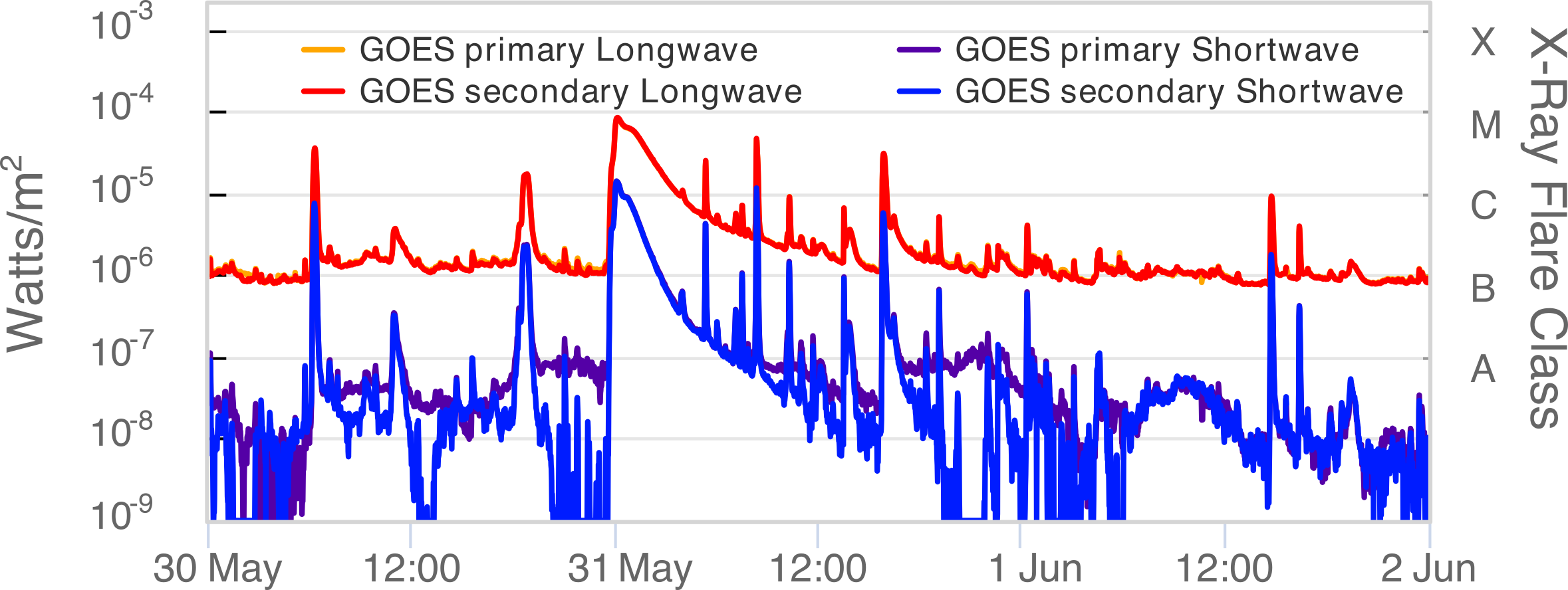}
    \caption{GOES scientific X-ray flux plot \cite{woods2024xrs1} shows the onset time of the 
    CME-associated flare to be 2025 May 30, 23:30 UTC.}
     \label{fig:goes}
\end{figure}

\subsection{The coronal mass ejection of 2025 May 31}\label{sec:halo-cme}

A bright, dense, fast-moving halo coronal mass ejection (CME) erupted from just northeast of the Sun's central meridian.  The associated flare onset time, marking the start of the event, was 2025 May 30 23:30 (Figure \ref{fig:goes}). The
corresponding halo CME front was tracked by SOHO/LASCO C2 and C3 \cite{brueckner_etal_1995}, crossing the 
the C2 field of view at approximately 2025 May 31 00:00 UT and rapidly expanding across both the C2 and C3 fields of view (Figure \ref{fig:lasco}). By 2025 May 31 05:00, 
the leading edge had an elongation $\varepsilon$ greater than $8^\circ$, and had therefore left the LASCO C3 field of view.  PUNCH 
imaged the propagating CME front for roughly 24 hours, starting at 01:00 UT.  Figure \ref{fig:punch-mosaics} shows PUNCH mosaic images at
six representative times, at four-hour intervals.  In these preliminary data, the front is fainter than the surrounding starfield (which is 
suppressed only by a factor of $\sim 30$ as described in the previous section), but is visible in the associated movie file and may be picked out even in individual 
frames.  At elongations beyond 20$^\circ$, the bright CME front is approximately 3\% to 10\% as 
bright as the variable ``background'' and is therefore difficult to pick out in still frames; its 
approximate location is marked on the figure.

\begin{figure}
    \centering
    \includegraphics[width=0.65\linewidth]{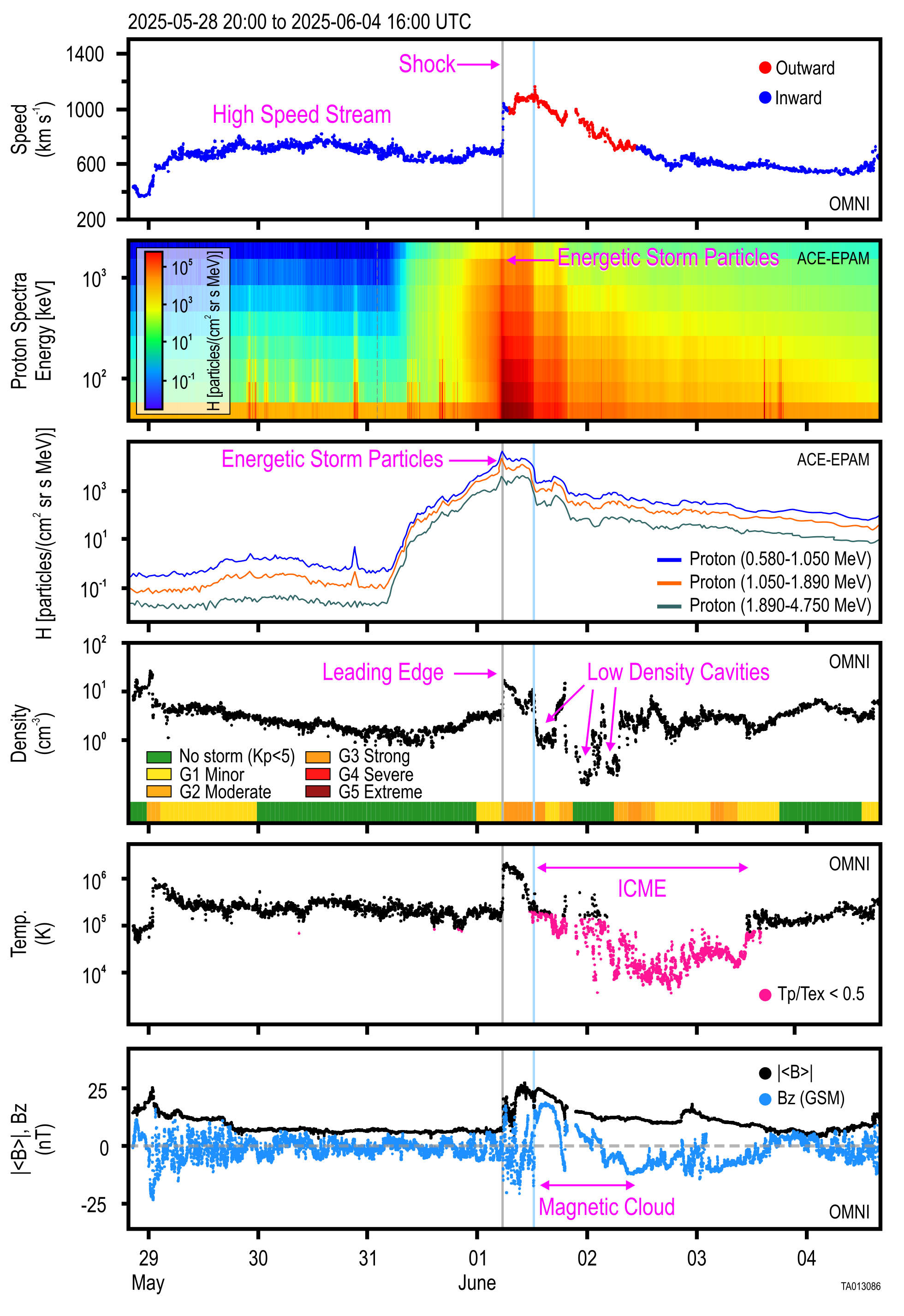}
    \caption{In situ observations of the halo CME observed by PUNCH on 2025 May 31 show: (a) the CME was running into an inward polarity high speed stream; (b) it was moving fast enough to form an interplanetary shock, which arrived at Earth at 2025 June 1 05:40 UTC (grey vertical line); (c)
    the event included typical features including an energetic storm particle (ESP) event, a magnetic cloud, and a low density central cavity. Top to bottom are time series of: OMNI solar wind proton speed; ACE EPAM solar energetic proton energy-time intensity spectrogram; time profiles of the SEP intensity for three selected energy passbands; the OMNI solar wind proton density; solar wind proton temperature; and the IMF field strength (black) and GSM $B_z$ (blue).}
    \label{fig:ace-figure}
\end{figure}

The May 31 CME impacted both Earth and space assets near L1 and can be characterized
more completely based on the in-situ signatures of its impact. Figure \ref{fig:ace-figure} shows the solar wind and IMF properties of the ICME from the OMNI data set, which is a combined collection of solar wind measurements from many spacecraft, propagated to Earth's bow shock for mutual comparison \cite{King2005,bluthner2026evaluating}. The 
energetic particle traces are directly from the ACE Electron Proton Alpha Monitor (EPAM) described by \citeauthor{Gold1998} \citeyear{Gold1998}; ACE is located at L1 some ~0.01 au upstream of Earth \cite{stone_etal1998}.  

The CME ran into/through an inward
polarity high speed stream (\ref{fig:ace-figure}), and had a clear magnetic cloud signature.  
The CME ejecta itself had 3
low density cavities, a clear CME signature indicated by a long continuous low-proton-temperature
interval where the temperature ratio $T_p / T_{ex}$ was less than 0.5
\cite{elliott_etal2016,richardson_cane2010}, and a distinct magnetic cloud signature consisting 
of a smooth magnetic field rotation \cite{burlaga_etal1981}. At first, the IMF $B_z$ was 
northward (positive) in the sheath, but rotated southward (negative) in the magnetic cloud, 
increasing its geoeffectiveness. The CME was geoeffective, reaching a G3 geomagnetic storm level
based on a maximum $K_p$ of 7+, and intense levels based on a minimum in SYM-H geomagnetic index of
-152 nT and minimum $D_{st}$ index of -119 nT. 

An interplanetary shock and sheath compression formed even though the ICME was running into a
fast wind stream with lower density than the slow solar wind. That compression forms the visible 
``CME front'' observed by PUNCH via light Thomson-scattered off the electrons entrained in the 
plasma.

The shock crossed over Earth on 2025-06-01 05:40 UTC (gray vertical line), with a 
peak trailing wind speed of 
1150~km/sec.  This compares to the average propagation speed, inferred from the CDAW flare time and 1~AU travel distance,
of 1360~km/sec, and is consistent with braking effects from the CME interacting with surrounding corona and solar wind en route \cite<e.g.,>[]{deforest_etal_2013a}.  

Enhanced levels of energetic particles are observed leading up to and peak at the shock, indicative of an Energetic Storm Particle (ESP) event. 
The shock showed compression ratios of  2.5, 1.7, and 4.0 respectively for the density, field strength and temperature. Compared to the statistical analysis of 278 interplanetary shocks in \citeA{moreland_etal2023}, the shock
temperature ratio is quite large, its density ratio is fairly large, and its field ratio is typical.

The 6 hours and 50 minutes compressed sheath interval between the shock on June 1, 2025 at 05:40 UTC and the time where the
density drops on June 1, 2025 at 12:30 UTC (light blue vertical line) has an average density of 9.7/$cm^3$, and a peak 
density of 19.0/$cm^3$. This is the portion of the CME that is most visible in Thomson-scattered light, and therefore is the
in-situ feature most closely corresponding with the ``CME front'' seen by PUNCH.

\begin{figure}
    \centering
    \includegraphics[width=0.85\linewidth]{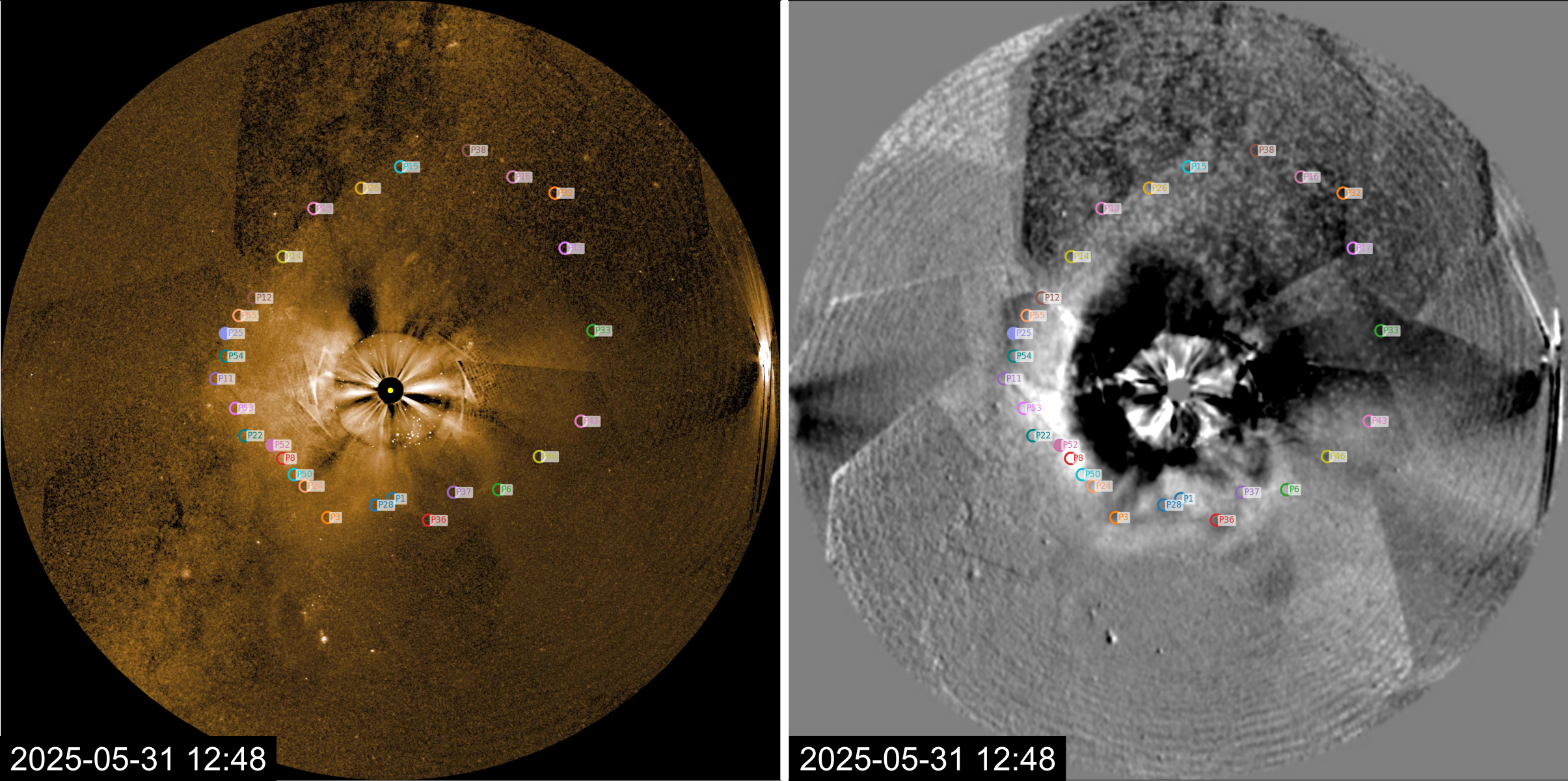}
    \caption{A single full-PUNCH-mosaic frame from 2025 May 31 reveals a halo CME propagating from Sun to Earth, using preliminary processing to eliminate F corona and reduce the brightness of the starfield.  Left: original image from 2025-05-31 12:48 UT.  Right: difference image from 2025-05-31 12:46 UT, with Gaussian blurring applied, highlights the front more than does the original image.  Points on the CME front are marked using the rotoscoping tool described in Section \ref{sec:rotoscoping}. The field of view subtends $94^\circ$ of arc. Note that, in the southeast quadrant, the event includes two identifiable ``leading edge'' features, the second of which (marked) overtakes the first as the event propagates.}
    \label{fig:sample-cme-image}
\end{figure}

\section{Methods} \label{sec:methods}

We conducted a  morphological study of the
CME as it propagated, using rotoscoping (Section \ref{sec:rotoscoping}) to identify the propagating front.
To demonstrate wide-field tracking, we compared the leading-edge position results to a 
simple semi-analytic shape model. (Section \ref{sec:modeling}).

\subsection{Rotoscoping} \label{sec:rotoscoping}

We tracked the CME front using the rotoscope method \cite[now often called ``point-and-click'']{fleischer1917} to extract a point cloud following the image-plane front 
of the CME, directly in the 
sequence of initial-release PUNCH mosaic frames.
We tried two different digital tools: a commercial rotoscoping software suite, and a custom-coded
tool that overcame some of the limits of the data and the commercial suite.

We began with the Rotoscope module of the Silhouette FX software \cite{silhouettefx}, which is 
designed for 
cinema and television
post-production. Two significant drawbacks led us to reject SilhouetteFX for our use.  First, it is licensed on a subscription basis, limiting reproduciblity of its results.  
Second, because it is optimized for 
hand-tracing features in cinematic data, SilhouetteFX did not allow real-time enhancements such as image differencing to improve detectability of the faint front. 

We switched to a custom-coded rotoscope tool, rotoscope.py \cite<available from Zenodo:>{deforest_etal_2026c}. We
included several software features that were useful for drawing out the front in places where it was faint or obscured: interactive blurring to reduce image noise; supplementary rendering of elongation-vs-time curves for each tracked point; and an interactive fixed-frame-difference feature that
allowed selecting a single frame and subtracting its value from arbitrary other frames, to remove slowly-varying backgrounds.  

Both tools allowed the operator to flip between frames, marking and tracing the propagating CME front with interpolation between 
key frames, and between key points within each frames, to form an evolving shape.  
We manually traversed the image sequence and adjusted the evolving location of each tie-point by clicking on the 
CME's leading edge at convenient key frames, verifying the location where possible in intermediate frames. The use 
of ad-hoc key frames and temporal spline interpolation allowed us to sample the front location in those frames where
it was most clear, interpolating across data artifacts in intermediate frames.  This permitted tracking 
the CME front despite variable background in the preliminary data, which resulted in irregular
and non-uniform sampling in time but captured well the front evolution as verified by comparing spline-interpolated 
front locations to the actual observed front in intermediate frames.  Further, the supplementary elongation plot allowed us
to reject false front identifications on the basis of inconsistency with other points adjacent in time and/or space.

After rotoscoping and interpolation we carried out further analysis in Python.  
We converted the ($X_{im},Y_{im}$)
pixel coordinates to solar position angle $\alpha$ and solar elongation angle $\varepsilon$ via
the azimuthal equidistant projection used by PUNCH: radius from image center is proportional 
to the solar elongation angle $\varepsilon$ on the sky; and
solar position angle $\alpha$ is the same on the image plane and in the sky 
\cite[Appendix A]{deforest2026}.  We
verified that the Python within-frame analytic spline interpolation matched the visual 
results from rotoscoping, then
used the calculated $\varepsilon$ and $\alpha$ angles for plotting (Section \ref{sec:results}).

\subsection{Geometric Modeling} \label{sec:modeling}

\begin{figure}
    \centering
    \includegraphics[width=5in]{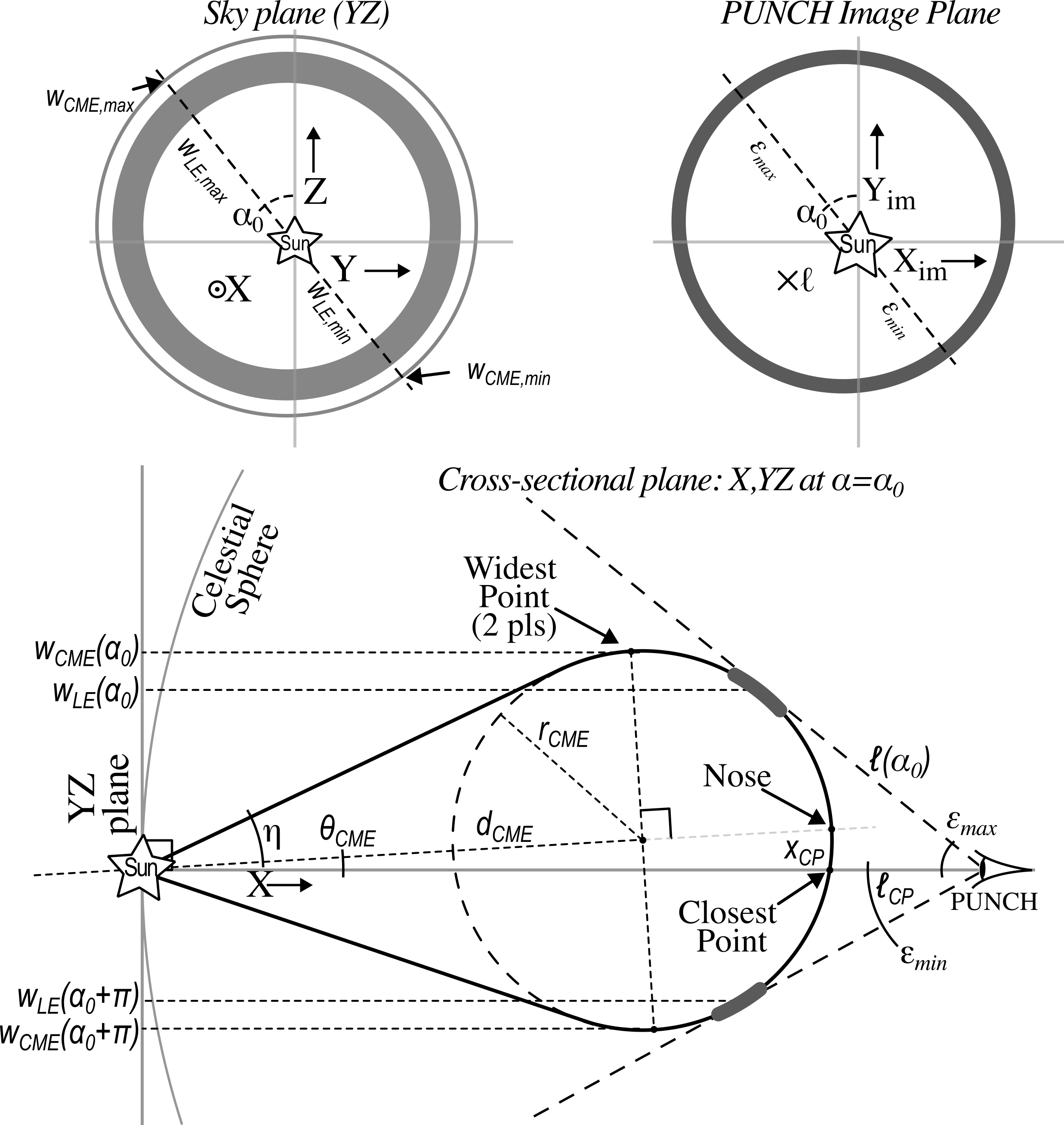}
    \caption{Cartoon diagram shows angles and geometry of PUNCH observation of a
    halo CME, along with relevant geometric quantities for the simplified ice-cream cone model described in the text.  
    The 3-D space between Sun and PUNCH is represented with the RTN coordinate system in which X extends from Sun to PUNCH, 
    Y is the cross product of solar North with X, and Z is the cross product of X with Y (upper left). PUNCH images are 
    Sun-centered azimuthal equidistant projections of the celestial sphere as seen from PUNCH, such that radius is 
    proportional to elongation angle $\varepsilon$ and solar position angle $\alpha$ is preserved (upper right).  
    The CME ``leading edge''seen from PUNCH is the locus of tangency between the CME front surface and lines of sight 
    extending from PUNCH to the celesial sphere (bottom).  Labeled quantities are more fully described in the main text.
    }
    \label{fig:ice-cream}
\end{figure}

Analytic treatments of CME geometry at wide angles abound and are applicable in different regimes of angle.  Some important landmarks in the development of these treatments include ``Fixed-$\Phi$''\cite{kahler_webb_2007},
``Harmonic Mean'' \cite{lugaz_etal2009}, ``self-similar expansion'' \cite{Davies_etal_2012}, and
parametrically evolving analytic \cite{tappin_howard_2009, wood_10b} approaches  including semi-empirical drag modeling \cite{rollett_etal_2016}.
We modeled the wide-field perspective of the CME using a crude analytic model inspired by, but
simpler than, the multi-parameter TH model from \citeauthor{tappin_howard_2009}: a ``ludicrous ice cream
cone'' structure similar to that developed by \citeA{xue_etal2005} and refined by 
\citeA{na2017development}.   
Keeping the simplicity of \citeauthor{xue_etal2005} we treat only a simple cone geometry (with 
apex at the Sun) and a 
spherical CME.  
The model
geometry is illustrated in Figure \ref{fig:ice-cream}, and is similar to that of 
\citeA{na2017development}.  The difference is that \citeauthor{na2017development} capped their
cone with a strict hemisphere, while we placed the spherical ``ice cream'' tangent to the cone.
This simplifies some of the geometrical calculations, and slightly changes the meaning of the 
cone angle although the models are clearly equivalent.  
Where relevant, we matched the names of geometrical quantities to the recommended PUNCH
nomenclature given by \citeA[their Appendix A]{deforest2026}.

The purpose of the model is to be as simple as possible while also representing the shifting 
perspective of the leading edge
in the wide-field PUNCH images: the observed leading edge is a point of tangency between 
lines of sight and the density 
shell at the front of the CME, and the spherical front permits the relationship between the leading edge and the 
3-D structure itself to vary as the CME approaches Earth.

The CME takes the appearance of an expanding,
slightly off-center halo in the PUNCH (X$_{im}$,Y$_{im}$) image plane, which
is an azimuthal equidistant projection of the celestial sphere, with 2-D origin at the Sun (upper right of Figure \ref{fig:ice-cream}).  
The CME ``leading
edge'' has a solar elongation angle $\varepsilon$ that varies as a function of position angle $\alpha$, with
a maximum of $\varepsilon_{max}$ at $\alpha_0$ and a minimum of $\varepsilon_{min}$ at $\alpha_0+\pi$.  The CME has scoop radius $r_{CME}$ and fixed cone half-angle $\eta$.  
The model expands with constant $\eta$ and steadily increasing $d_{CME}(t)$. The direction of
propagation is slightly off center: it is canted by an angle $\theta_{CME}$ from the 
PUNCH-Sun line, with solar position angle $\alpha_0$.  In the three-dimensional Cartesian RTN system,
the CME maximum sky plane offset $W_{CME}(\alpha)$ and leading edge sky plane offset $W_{LE}(\alpha)$ 
are projected on to the YZ plane perpendicular to the Sun-PUNCH line. For all values of $\alpha$, 
$W_{CME}(\alpha)>W_{LE}(\alpha)$, because the ``leading edge'' is the locus of tangency between 
the CME front surface and lines of sight extending from PUNCH to the celestial sphere.  

We considered the CME to have a fixed half-cone angle of $\eta$, and to expand self-similarly at uniform 
speed so that $r_{CME}\propto$ $d_{CME} \propto X_{CME}\propto t-t_0$, where $r_{CME}$ is the radius of the spherical portion of 
the geometry 
(the ``scoop''), $d_{CME}$ is the distance of the center of the scoop from the Sun, $X_{CME}$ is the X coordinate of the center of the scoop, $t$ is absolute time, and 
$t_0$ is the time of CME launch from the Sun.  In fitting the model across time, we defined $v_{CME}$ as $d/dt(d_{CME})$, although we note that the time derivative of any of the mutually-proportional distances would have sufficed. 

For small elongations $\varepsilon < 15^\circ$, where the small-angle approximation applies, the curvature 
of the trigonometric functions is negligible, so that $\sin 
\varepsilon \sim \varepsilon$, and $\cos \varepsilon \sim 1$.  In that condition, with constant-speed 
radial (i.e. constant-velocity) CME motion and self-similar expansion, $\varepsilon \propto \eta\,v_{CME}\,t$, and it is not possible to 
determine simultaneously $\eta$ and $v_{CME}$ from a single image sequence \cite<e.g.,>{tappin_howard_2009,DeForest_etal2016b}.  
That is important, because determining $v_{CME}$ is critical for the problem of space weather forecasting, in 
which $\theta_{CME}$, the distance parameter $\ell_{CP}(t)$, and the impact time $t_{\ell=0}$ are desired 
and the principal input is a time series of halo shape measurements of the 2-D ``leading edge'', $\varepsilon_{LE}(\alpha,t)$.

\begin{figure}
    \centering
    \includegraphics[width=0.96\linewidth]{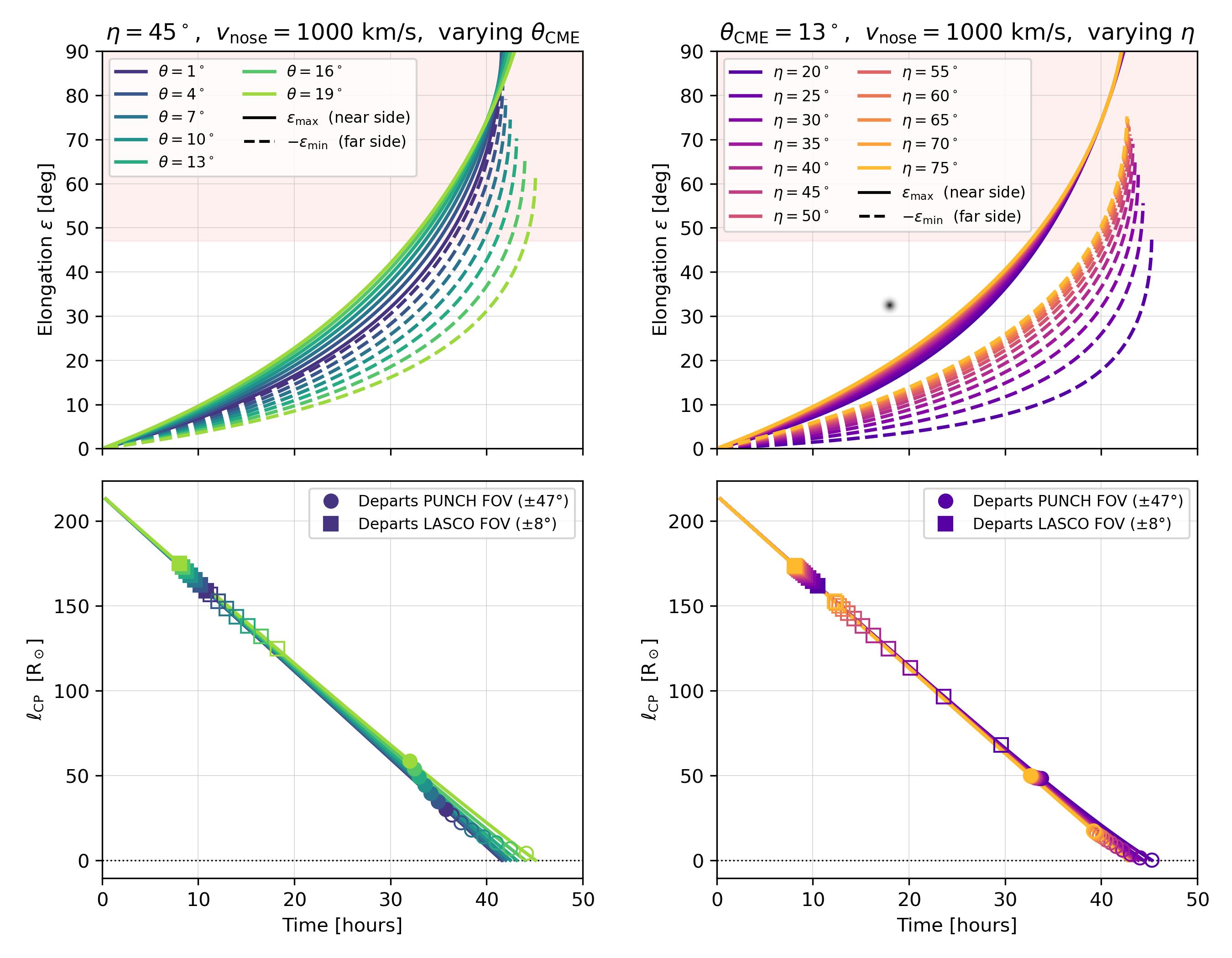}
    \caption{Sample model propagation curves show the behavior of notional CMEs with the ice-cream cone model
    of Figure \ref{fig:ice-cream}.  The left column (warm colors) shows behavior of an ensemble of halo
    CMEs, launched slightly off center (i.e., with fixed $\theta_{CME}=13^\circ$), with varying $\eta$ angle (half-cone width).  The 
    right column (cool colors) shows behavior of a similar ensemble, with fixed $\eta$ and varying impact angle 
    $\theta_{CME}$. One case ($\theta_{CME}=13^\circ$, $\eta=45^\circ$) appears in both columns.  The $\eta$--$v$ measurement degeneracy
    is broken by the varying curvature of the different traces at high solar elongation ($\varepsilon > 20^\circ$), as is seen in the varying degree of curvature as $\eta$ increases.
    In both columns,
    the top plot shows elongation angle $\varepsilon$ vs time, with solid lines showing $\varepsilon_{max}$ in the
    direction of CME offset ($\alpha = \alpha_0$) and 
    dashed lines showing $\varepsilon_{min}$ in the opposite direction ($\alpha = \alpha_0 + \pi$).  Also in both cases, 
    the bottom plot shows distance from the CME
    impacting front, to Earth, over the same time range.  The area outside the PUNCH FOV is shaded light red in the top plots.}
    \label{fig:model-scans}
\end{figure}

As the CME propagates, $\varepsilon$ expands beyond $15^\circ$, trigonometric nonlinearity 
becomes important, and it becomes in-principle possible to infer $\eta$ and $v_{CME}$ simultaneously,
among other geometric parameters, by tracking the time dependent 2-D geometry of a CME leading edge
\cite<e.g.,>[]{howard_tappin_2009,davies_etal_2009}.  The method is subject to further degeneracy or near-degeneracy between the curvature of the CME leading face 
and its location in 3-D space; and this effect is exacerbated by real-world effects in which 
energetic CMEs interact with their surroundings, both braking/drag by mass accretion on 
the leading face and also changing their shape in response to the same accretion.  
So simplified models like the constant-velocity ice-cream cone model are limited at best.  
Indeed, PUNCH was specifically designed to break those  
degeneracies using polarization of
light; and that work is anticipated for 
future analyses once polarimetric data become available.

Figure \ref{fig:model-scans} shows the behavior of the ice-cream cone model in the PUNCH field of view.  The upper two
plots show variation of the leading-edge $\varepsilon$ over time for a family of models with 
similar $\theta_{CME}$ and varying
$\eta$ (left) and a family of models with similar $\eta$ and varying $\theta_{CME}$ (right) (all quantities 
from Figure \ref{fig:ice-cream}).  The lower two panels show the progression of the closest-to-Earth point on the 
spherical portion of the ice-cream cone, for the same families of model.  The symbols on the lower plots show the times
at which the modeled CME front leaves the LASCO (square) and PUNCH (circle) field of view for each model from the family.  The solid symbols show the 
maximum-$\varepsilon$ exit (outermost part of the halo); and the outlined symbols show the minimum-$\varepsilon$ exit 
(final exit of the halo from the corresponding field of view).  The degree of curvature in the two top plots breaks
the degeneracy (between $\eta$ and $v_{CME}$) that is present in narrow-field measurements.  This is not possible with
narrow fields of view; for example, the SOHO/LASCO C3 coronagraph extends to less than $8^\circ$ from the Sun, still
well inside the narrow-angle regime.  In general, the sharpness of the bend in elongation-vs-time plots (as in Figure \ref{fig:model-scans}) indicates the width of $\eta$, with shallower bends corresponding to wider values of $\eta$; and the asymmetry of the bend indicates the value of $\theta$.  Even in the widest case plotted ($\eta=75^\circ$, corresponding to a $150^\circ$ wide CME), the shallow (worse) side of the CME still exhibits a factor-of-two shift in slope across the marked WFI field, breaking the speed-width degeneracy for all plausible cases (at least with this simplified geometric model).

For simplicity, we used a four-parameter
fit of the extended family of models, varying $\theta_{CME}$ $\alpha_0$, $\eta$, and $v_{CME}$, to fit our 
leading edge data across time, using least-squares optimization and the ``amoeba'' simplex fitter \cite<e.g.,>{Nelder_Mead_1965,press1992book}.  We found
that $\eta$ tended to converge to its theoretical maximum 
of $90^\circ$ (``bubble'' model with the CME front coinciding 
with the 
Thomson surface at impact).  We judged that this was not plausible, and attributed the 
anomaly to the small-angle $\eta$--$v_{CME}$ degeneracy, 
since most points
in the dataset are either in or close to the small-angle regime and only a few are in the wide-angle regime where 
the ``knee'' is located in Figure \ref{fig:model-scans}.  
In the spirit of using the simplest possible model, we 
chose a large-but-not-maximal value of $\eta=60^\circ$ and reverted to a 
three-parameter fit of 
$\theta_{CME}$, $\alpha_0$, and $v_{CME}$. All subsequent fits used this arbitrary value
for $\eta$.  As a sanity check, the CCMC DONKI catalog reports a modeled $\eta$ of 
47$^\circ$ for this CME \cite{donki_cmeanalysis}, which is different from the 
60$^\circ$ we chose but not drastically so given the $\eta$--$v_{CME}$ degneracy at small angles.  The effect of $\eta$ selection is further discussed in Section \ref{sec:testing}, below.

We found that setting $t_0$ to the known flare onset time yielded 
poor fits, with either early or late halo geometry suffering when the other end of the time 
range was
weighted more heavily.  That is to say, we tested the quality of the model by weighting either early or late times
more heavily, and checking the quality of the geometric fit either late or early, correspondingly. We found that the
model could not capture both early and late expansion rates of the halo at the same time under that condition. 
We attributed this effect to the known deceleration of very fast CME fronts; this one was indeed very fast, 
with an observed average
speed of 1300~km/sec across 1~AU from flare onset time to impact with Earth.  Fits weighted to early front positions tended to yield speeds 
closer to 1800~km/sec, and fits weighted to late front positions with a fixed start time tended to converge to the average speed.  Deceleration is typical of fast, strong CMEs, like this one, that are associated with
large flare events \cite<e.g.,>{tappin_howard_2009,rollett_etal_2016}.  

Rather than add a further
deceleration parameter to the fits, we added an ad-hoc offset in the start time, modeling $t_0$ 
as if the 
CME started 8 hours earlier than the flare event.  The value of 8 hours was chosen as a
round number 
based on an \textit{a priori} estimate of a typical $\sim$~30\% slowdown between early propagation and the Sun-to-Earth average speed in the 
heliosphere, from cursory inspection of several fast halo events in the CDAW catalog \cite{gopalswamy2010}, and was not adjusted for this particular event.  We took, as an \textit{ansatz}, that the $t_0$ formulation can account for the deceleration well enough to characterize the CME's behavior far from the Sun once the strongest effects of CME braking are over.  We then validated the use of the nearly-arbitrary $t_0$ via a parametric study (described in Section \ref{sec:testing}).

\section{Results} \label{sec:results}

We analyzed the data points collected by rotoscoping, first assessing the front behavior 
(Section \ref{sec:leading-edge-results}), then comparing to in-situ data 
(Section \ref{sec:data}) and the ice-cream cone model (Section 
\ref{sec:model-results}), before evaluating their utility in forecasting 
(Sections \ref{sec:forecasts} and \ref{sec:testing}).

\begin{figure}
    \centering
    \includegraphics[width=0.75\linewidth]{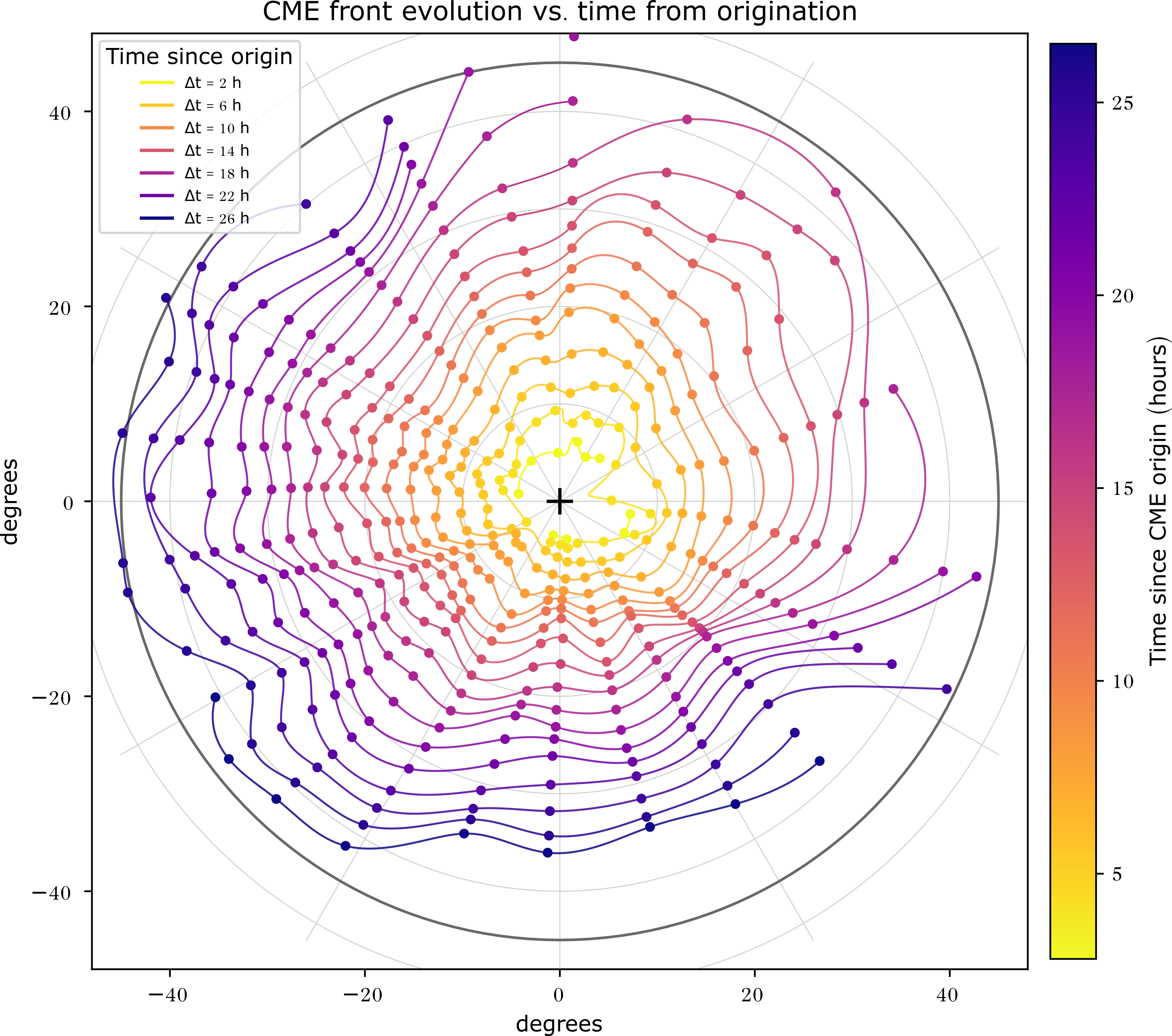}
    \caption{Evolution of the CME front in the PUNCH field of view, as captured with our rotoscoping tool, shows slight asymmetry (also visible in Figure \ref{fig:sample-cme-image}), with faster image-plane propagation in the
    northwest and slower in the southeast.  The CME completely left the field of view approximately 27 hours after the flare onset time of 2025 May 30 23:30.  Marked points 
    indicate individual tracked points around the front, and are spline-interpolated in time from the
    individual rotoscoped locations and frames to the uniform
    cadence in the figure.  The
    curves are spline-interpolated between the tracked points, to approximate the overall front.  The black circle represents $\varepsilon=45^\circ$ from the Sun, the edge of the nominal WFI field of view.  The Sun's position is marked
    with a ``+'' at the origin.  The figure is oriented so that solar north is up.  Data were available to
    approximately $47^\circ$. }
    \label{fig:front-evolution}
\end{figure}

\begin{figure}
    \centering
    \includegraphics[width=1.0\linewidth]{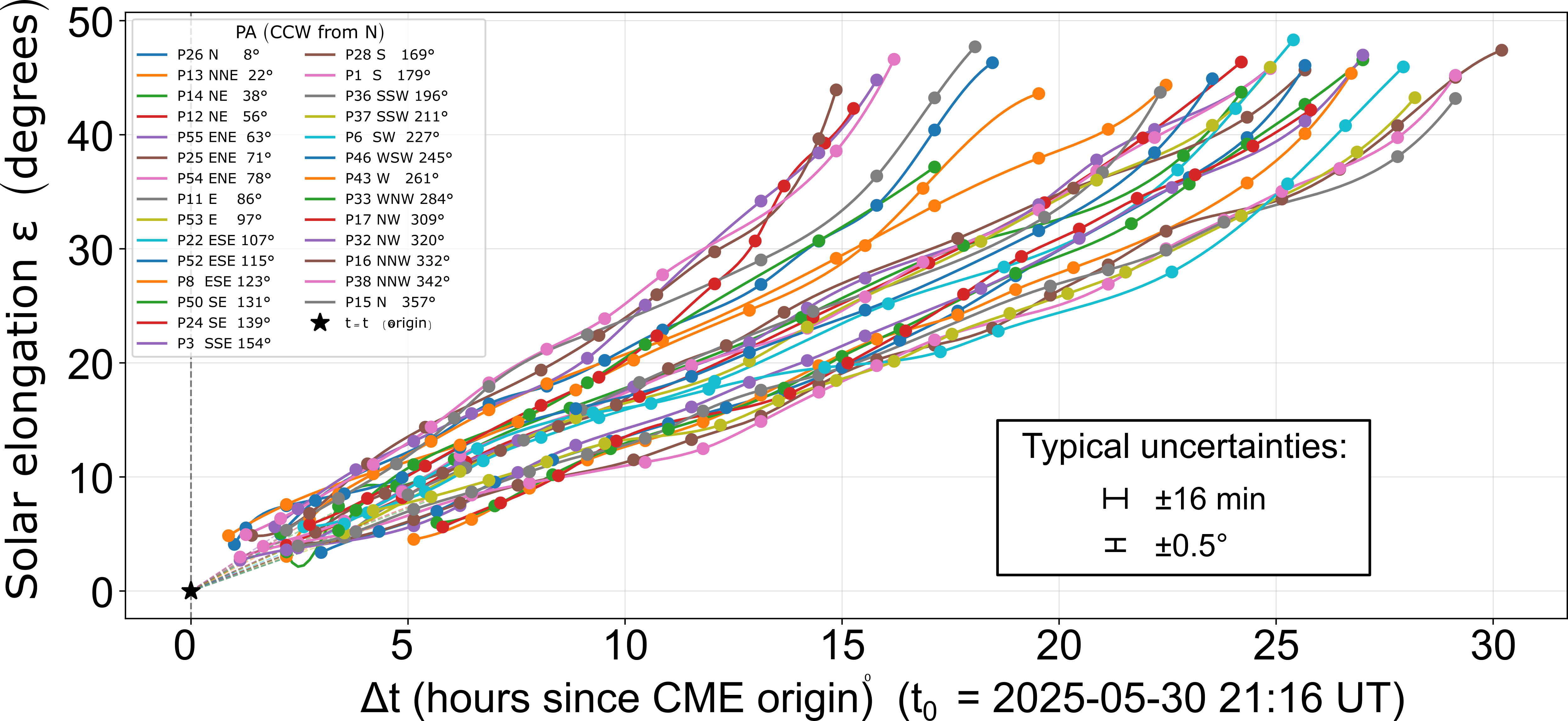}
    \caption{Time plots of each point show how $\varepsilon$ of the front varies with time, for all samples. The
    points are most consistent with a calculated start time of 2025-05-30 21:16 UT based on
    constant-velocity extrapolation of the trajectory plots using all points with $\varepsilon < 15^\circ$; this is moderately consistent with the observed flare time of 2025-05-30 23:30 UT. Note the general trend of increasing angular speed at high 
    elongations (also visible in Figure \ref{fig:front-evolution} as wider spacing of the front curves at higher elongations, and consistent with Figure \ref{fig:model-scans}).  Placement error is roughly $\pm 1^\circ$ in $\varepsilon$ and $\pm 16$ min in $\Delta t$. 
    }
    \label{fig:evolution-plot}
\end{figure}
\subsection{CME Leading Edge Analysis} \label{sec:leading-edge-results}

We rotoscoped the CME front using point propagation along approximately radial loci in frames that captured the CME motion (``key frames''), with spline interpolation on the
time axis.  Error values were difficult to estimate a priori given the multiple, and highly variable, error sources
in the preliminary data.  We estimate radial positioning error to be $\pm 0.5^\circ$ (approximately 10 2$\times$2 binned PUNCH pixels).
We estimate timing error to be $\pm 16$ min (up to two samples in either direction), due mainly to irregular 
sampling as the WFI constellation had not yet
achieved equal spacing around its orbit.

The rotoscoping data yielded a 3-D point cloud with each point located in $X_{im}$, $Y_{im}$, and $t$.  
The points were organized into tracks represented by numerical functions, with $X_{im}$ and $Y_{im}$ being dependent 
variables and 
$t$ being the independent variable.  We identified the front visually, approximately every 10th 
frame (``key frames''),
and followed the
spline-interpolated location of each point in intermediate frames to check that it indeed followed the visual front
between key frames.  The key frames were generally different between different tracks, as we selected them for 
high visibility of the front at that particular position angle $\alpha$.  These spline-interpolated ($X_{im}, Y_{im}$) 
locations were used to create the marked points in Figure \ref{fig:front-evolution}, showing the tracked front every 40 minutes (every fifth PUNCH mosaic) throughout its travel.

Figure \ref{fig:front-evolution} includes curves connecting the individual point locations at each time.  These curves are spline-interpolated in space, between the tracked points, to visually represent the evolution of the CME front over time.

Being able to use ad hoc differencing and smoothing helped reveal features that were not visible in the direct
images.  Initial tracks with SilhouetteFX showed anomalous acceleration at approximately $15^\circ$ elongation, 
in the southeast portion of the field (lower left).  Using differencing, we were able to see that the initial 
``bright front'' comprised two separate eruptions at slightly different times, with the earlier front being overtaken
by the halo that ultimately impacted Earth.  This is visible in Figure \ref{fig:sample-cme-image} as a bright fringe
outside the marked points.  Following the later front yielded smooth velocity contours with no sudden 
anomalous acceleration. 

Both in the direct images and in the difference images, we noticed that the
bright front was quite sharp, and the spatial error bar in Figure 
\ref{fig:front-evolution}
is therefore conservative. We refrain from detailed physical interpretation of the 
front's nature, though we note that (a) bright features in images formed from Thomson-scattered light correspond to density 
features in the solar wind, and (b) prior studies 
\cite<e.g.,>{ontiveros_vourlidas_2009} have identified imaged, sharp-edged bright 
fronts with shocks.

A direct visual result from Figure \ref{fig:front-evolution} is that the 
developing off-center trajectory of the halo is immediately clear from the data.
The large-scale asymmetry is a 
direct indicator of the $\theta_{CME}$ and $\alpha_0$ parameters, which describe
whether Earth interaction is expected to be head-on, glancing, or missing. 
The evolution 
of the degree and direction of asymmetry indicate whether the CME is 
deflected as it interacts with 
the surrounding 
interplanetary medium. In this case, the uniformity of the visual center 
immediately indicates that the CME was propagating approximately radially.

Figure \ref{fig:evolution-plot} shows the evolution in time of each tracked point.  The observed 
radial speed close to
the Sun (within $15^\circ$) varies between 1.4 and 2.2 degrees per hour; this is equivalent, 
in the small-angle approximation commonly used to analyze coronagraph images, to 1000 to 1600~km/sec 
in the plane of the sky, and is typical of the 
fastest halo CMEs observed
with LASCO \cite<e.g.,>{StCyr2000,gopalswamy2010}. The general trend is of gradually increasing 
angular speed as the front extends farther into the heliosphere and away from the small-angle regime, consistent with Figure 
\ref{fig:model-scans} and with earlier work \cite<e.g.,>{michalek2003,webb_etal2009b}.  

An immediate visual result from both Figure \ref{fig:front-evolution} and Figure \ref{fig:evolution-plot} is that the CME is not particularly
spherical or smooth.  While the front is approximately self-similar over time, small loci within the CME move
faster or slower than the average, and the overall shape is quite variable on angular scales of a few degrees 
relative to Sun center.  The apparent speed of the leading edge shifts, in some cases abruptly, in both 
space and time, indicating the importance of multiple disconnected structures within the body of the CME.  We attribute the abrupt shifts to perspective effects from substructures overtaking one another in the image plane, as the perspective shifts.  In some, but not
all, cases, the overtaking structure can be identified in the preliminary image data (such as the aforementioned double-front in the southeast, visible in Figure \ref{fig:sample-cme-image}). Determining the balance will require using the lower-noise images available with subsequent reprocessing.

To sanity-check the tracking and identify whether the simplified geometric/analytic model was 
even applicable, we estimated the start time of the CME by extrapolating the earliest 8 hours of each 
track backward in time with linear extrapolation, to identify how well and when they converged to a point, and to 
compare this ``best-fit launch time'' with the known launch time.  
The extrapolated launch time is 2025-05-30 21:16 UT, which is similar to (though slightly earlier than) the 
flare onset time of 23:30 UT. 
We attribute the discrepancy to braking in the outer corona as the rapid 
CME impacted and accreted overlying material.  This justifies the decision in 
Section \ref{sec:modeling} to use
a temporal offset in subsequent ice-cream cone model fits.

\subsection{Comparison to model} \label{sec:model-results}

\begin{figure}
    \centering
    \includegraphics[width=0.75\linewidth]{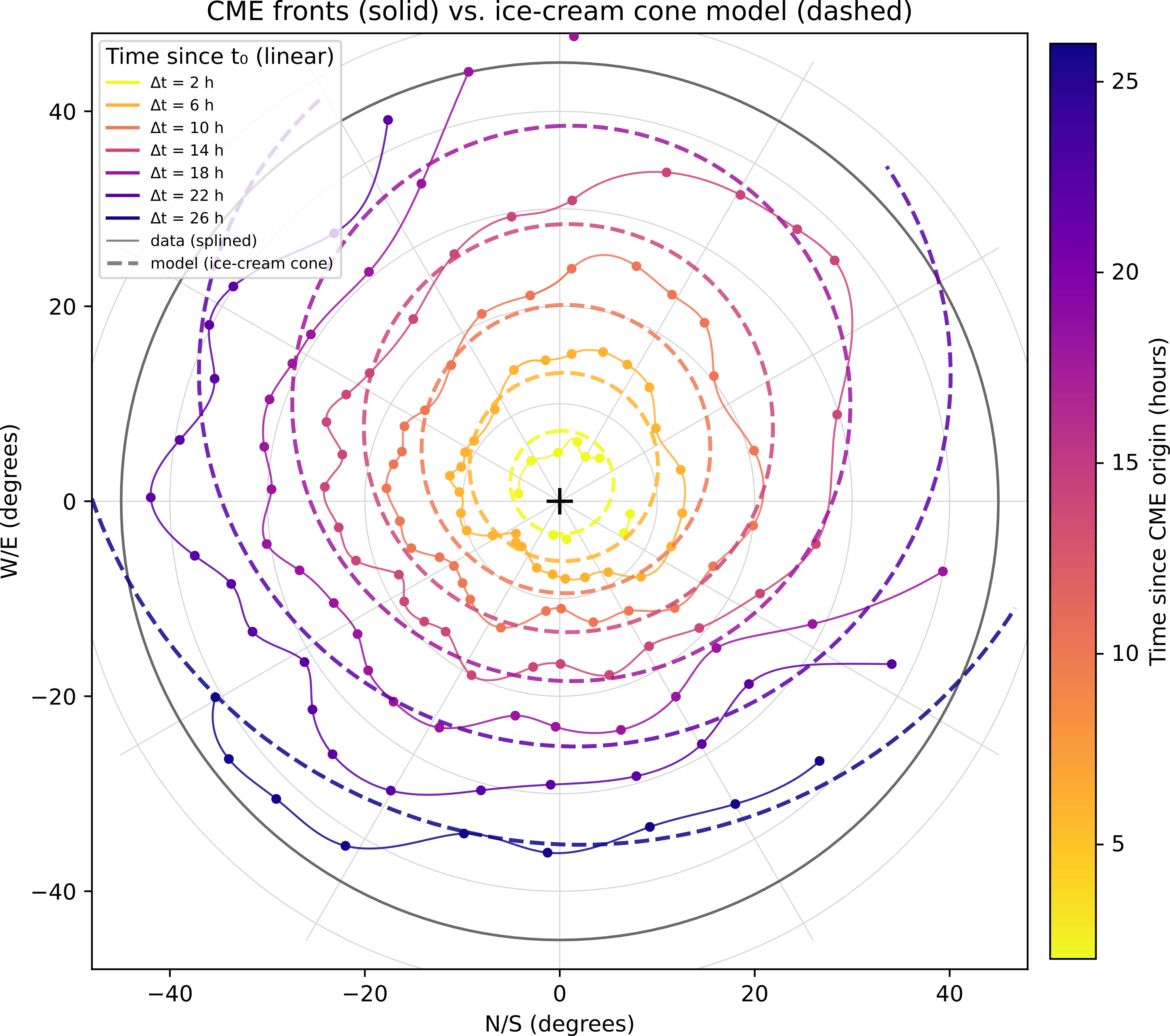}
    \caption{Contours of the fitted CME front, together with the best-fit ice-cream-cone model, show that the fit captures the off-axis nature of the CME and the gradual acceleration of apparent motion in $\varepsilon$ across the expansion of the CME. Fit residuals are dominated by two identifiable sources that are visible in the figure: (1) the shape of the 
    CME is more complex than a sphere; (2) the actual velocity is not constant, leading to a systematic trend across time in the offset between data points and simultaneous fitted contours, despite a zeroth-order correction via forced $t_0$.}
    \label{fig:ice-cream-fit}
\end{figure}

We parameterized the ice-cream cone geometry as described in Section \ref{sec:modeling} and fit it
to the extracted point cloud in Section \ref{sec:leading-edge-results}.  The resulting model
parameters are given in Table \ref{tab:fit-results}.  The modeled front shape is given in Figure \ref{fig:ice-cream-fit}.

\begin{table}[]
    \centering
    \begin{tabular}{c|c|c}
       Parameter  &  Value  & Comment\\
       \hline
       $t_0$  & 2025-May-30 15:30 & Forced to $t_{flare}$ - 8~h  \\
       $v_{nose}$ & 1300 km/s & $6.7 r_{\odot}/hr$ \\ 
       $\theta_{CME}$ & $18.6^\circ$ & near, but not on, Earth-Sun line\\
       $\alpha_0$ & $-7^\circ$ & NNW (consistent with point cloud) \\
       $\eta$ & $60^\circ$ & Forced to $60^\circ$ \\
    \end{tabular}
    \caption{Ice-cream cone model fit parameters (RMS residual=$3^\circ$ in $\varepsilon$)}
    \label{tab:fit-results}
\end{table}

The ludicrously simplified ice-cream cone model (with two of the initial five parameters forced to fixed values) does capture the slightly off-axis nature and gradually increasing angular speed of the CME,
as is apparent from the general agreement between the circular dashed contours and correspondingly colored data splines in 
Figure \ref{fig:ice-cream-fit}.  Shape residuals exist in two major forms: higher spatial frequencies (``bumps and valleys'' 
in each contour) that are real deformations of the front; and lower spatial frequencies, i.e., the overall (and systematically growing) shape mismatches between
the front and the model, especially in late contours.

That the front deforms and develops complex shape should not be surprising: with $\eta=60^\circ$, the full-width lateral expansion rate of 
a 1300~km/s CME is well above the typical solar wind Alfvén speed of order 100~km/s.  At angular offsets of order $10^\circ$ around a halo, the differential expansion rate is
comparable to the Alfvén speed, so that the entire CME ``leading edge'' is best 
considered as a set of causally disconnected plasma systems flying in loose
formation and interacting with their separate environments, rather than as a coherent
whole.  This effect may be exacerbated by spatial variability of the solar wind 
throughout the inner heliosphere, as different parts of the CME encounter different
solar wind conditions enroute.
Understanding the evolution of CME interior structure in more detail is a 
science objective of the PUNCH mission and will be addressed by future work 
\cite<e.g.,>{deforest2026,malanushenko_etal_2025}.

A more subtle residual is visible in Figure \ref{fig:ice-cream-fit}: there is a systematic temporal shift between
the size of the modeled circular fronts and the splined tracking results, with the innermost model being too large, 
intermediate models being notably too small, and the outermost two models matching reasonably well.  This is a combination
of two effects: the $\eta$ curvature effect visible in the top panels of Figure \ref{fig:model-scans}; and the 
fixed-speed model we applied, which does not capture the braking of the CME by accretion of surrounding slower-moving plasma. 
Separating these effects is in principle possible, but is beyond the scope of this exploratory work.  We
content ourselves with pointing out that the model, while reasonable, is quite flawed compared to contemporary geometric 
models used to forecast CME arrival times \cite<e.g.,>{Millward_etal2024}.

\subsection{Arrival time forecast simulation} \label{sec:forecasts}

\begin{figure}
    \centering
    \includegraphics[width=0.8\linewidth]{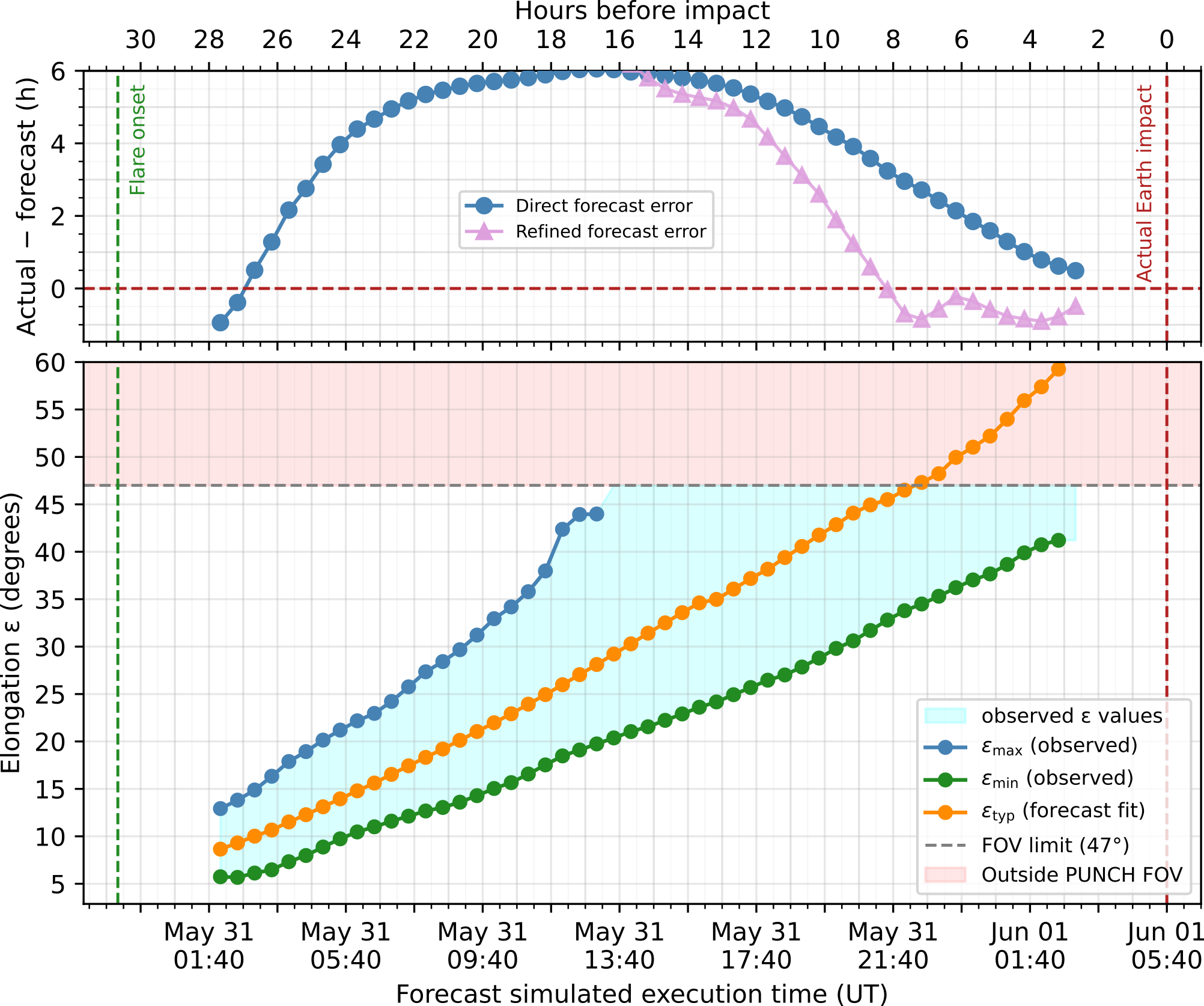}
    \caption{Retroforecast results using the ice-cream cone model show that even a crude analytic model can converge to create an accurate arrival-time forecast.  Half-hourly arrival-time forecasts using three-parameter, constant-velocity ice-cream cone models as in Figure \ref{fig:ice-cream-fit}, fitted using only data received before each forecast, diverge early in the CME propagation interval. As the CME approaches, forecast arrival time
    converges to within 30~$\pm 15$ minutes of the actual time of impact, shown in Figure \ref{fig:ace-figure}; the final forecast value is just 20 minutes off.  Refining the forecast using multi-fit extrapolation yields stable forecasts within $30~\pm15$ minutes of the actual impact time.  The lower panel shows the maximum ($\alpha=\alpha_0$), minimum ($\alpha=\alpha_0+\pi$), and 
    modeled intermediate ($\alpha=\alpha_0+\pi/2$) elongations of the tracked halo front.  See text for discussion. }
    \label{fig:forecasts}
\end{figure}

As a proof-of-concept for geometric forecasting using $360^\circ$ wide-field imaging in the image plane, we 
used the same three-parameter ice-cream cone model as a simple ``forecast'' tool.  This exercise simulates 
the use of a dedicated heliospheric imaging system with half-hour full-field cadence (like PUNCH)
and comparable data latency (unlike PUNCH).
PUNCH itself is not operated for low latency operation: the ground-pass cadence is approximately 20 hours, 
driven solely by the 
need to downlink images for scientific analysis (though the flight assets as-built could support more than 
one pass per hour continuously if needed).  
The exercise is useful as a proof-of-concept for the potential effects of near-real-time $360^\circ$ wide-field imaging on future 
forecasting efforts.

At each forecast time $t_{fc}$, we generated a forecast by fitting the analytic model only from data collected earlier 
than $t_{fc}$ using the three-parameter fit described above.  We then extrapolated in time to find the collision time of that
model with Earth (i.e. $t_1$ such that $\ell_{cp}(t_1)=0)$, given the fitted $v_{CME}$. ($\ell_{cp}$, the
distance to Earth, is shown in Figure \ref{fig:ice-cream}). 

To overcome the limitations of the constant-velocity model, in each forecast fit we weighted more ``recent'' 
data more strongly than ``older'' data, by multiplying the error term from each data point, by a Gaussian 
function in the difference between $t_{fc}$ and the acquisition
time of each datapoint. The half-width of the Gaussian was chosen to be six hours.  This had the effect of
most heavily weighting recent data points for each forecast, and smoothly dropping the significance of earlier
data to near zero.  The six-hour time constant was chosen to ensure that a significant interval of CME propagation was included 
in each fit, while ignoring the earliest points as the CME approached Earth; we made no attempt to optimize the 
time constant.  

We plotted the difference between each forecasted arrival time and the actual known arrival time
at Earth, vs. $t_{fc}$, to show how arrival-time forecasts might be updated as new data arrive.  The result is 
plotted in the top portion of Figure \ref{fig:forecasts}.  

The shape of the curve bears discussion.  The effect of CME braking/drag
is known and significant.  Letting $t_0$ be the same as the actual flare onset time would therefore yield very early
forecasts based on expansion data through the corona, without taking drag into account.  We confirmed this by setting 
$t_0=t_{flare}$ (not shown in Figure \ref{fig:forecasts}) and initial fits predicted arrival in as little as 15 hours 
from the flare itself.  Setting $t_0=t_{flare}-8$~hr (as in Figure \ref{fig:forecasts}) partially compensated for this 
effect.  However, braking continued throughout the propagation, so that the constant-velocity fits again trended early 
at intermediate distances.

The lower panel of Figure \ref{fig:forecasts} shows the maximum ($\alpha=\alpha_0$), minimum ($\alpha=\alpha_0+\pi$), and 
typical ($\alpha=\alpha_0+\pi/2$) elongation angle of the CME as it propagated, for comparison with the ensemble of curves
in Figure \ref{fig:model-scans}.  The typical elongation $\varepsilon_{typ}$ 
is shown from the fitted model, rather than the data, because (a) it agreed well with that derived from the data, where the whole 
halo was visible, and (b) as data points left the field of view the average $\varepsilon$ across data points became skewed.
As might be expected, the $\varepsilon_{min}$ curve tracks the bottom of the ensemble in Figure \ref{fig:evolution-plot}, 
and the the $\varepsilon_{max}$ curve tracks the top of the same ensemble.  The cyan region shows the range of elongations 
$\varepsilon$ at which the CME was visible (across different values of position angle $\alpha$); and the red region shows areas
outside the PUNCH field of view.  The $\varepsilon_{min}$ curve is quite flat, corresponding to the broadest values of $\eta$
in Figure \ref{fig:model-scans}, and the $\varepsilon_{max}$ curve bends up slightly, moderately consistent with the 
large-but-not-maximal value of $\eta=60^\circ$ that we chose.  (Note that the points in Figure \ref{fig:forecasts} represent 
individual forecasts, not individual
data points; a glance at Figure \ref{fig:evolution-plot} reveals that the upturn is significant, occurring across several 
position angles and for several sample points on each track.)

The main result from Figure \ref{fig:forecasts} is that, even with this very simple three parameter
model -- with known flaws including the simplistic CME-as-a-sphere geometry, pre-set $\eta$, and 
constant-velocity propagation -- collecting wide-field data permitted continuous refinement of the arrival time, 
with the final forecast predicting impact within 20 minutes of the 
actual arrival.  This continuous improvement is due to 
two complementary 
effects: (1) as time progresses and the CME crosses the inner solar system,
the average speed as derived from the model better approximates the actual
average speed; and (2) as the remaining distance decreases, so also the effect of  error in that derived speed decreases.
Effects which would render the forecast essentially useless for extrapolation from the coronagraphic field, 
became insignificant as the CME approached Earth and closer tracking data became available.

Even without a sophisticated propagation model, it is possible to improve the forecast further.  
At each forecast time, we linearized the relationship between predicted 
arrival time and the simulated forecast-execution time using only the current and prior forecasts, and found the root: the time at which 
predicted arrival time was itself predicted by that linearized model to coincide with the simulated 
forecast-execution time.  To reduce jitter in this projected forecast-of-forecasts, we averaged across the last four
projected forecasts.  This produced the results in the ``refined forecast error'' trace in
Figure \ref{fig:forecasts}.  During the final stages of approach, this refined forecast settles, 
and remains stable, within $30 \pm 15$ min of the 
actual Earth impact.    

We attribute this surprising stability of the final forecasts to the 
effectiveness of Taylor expansion in approximating smooth functions. Errors in $\eta$ or $t_0$ yield errors in the direct forecast time, due to mismatch between
the simplified model and the physics of propagation.  This class of error 
converges smoothly to zero as the arrival time approaches (blue trace at top of
Figure \ref{fig:forecasts}), so that linear extrapolation compensates for most
of the expected ``$\eta$ effect'' and ``$t_0$ effect'' in the final hours of 
propagation.

\subsection{Testing the Forecasts}\label{sec:testing}

The simplified ice-cream cone model in Section \ref{sec:forecasts} is remarkably effective, but it
has two forced parameters: $\eta$ and $t_0$.  One concern is that the parameters themselves
could in principle make the model work from sheer luck.  To eliminate that hypothesis,
we tested the same forecast method in Section \ref{sec:forecasts} with ensemble models to 
observe how the forecasts might vary with the selected parameters. 

\begin{figure}
    \centering
    \includegraphics[width=0.9\linewidth]{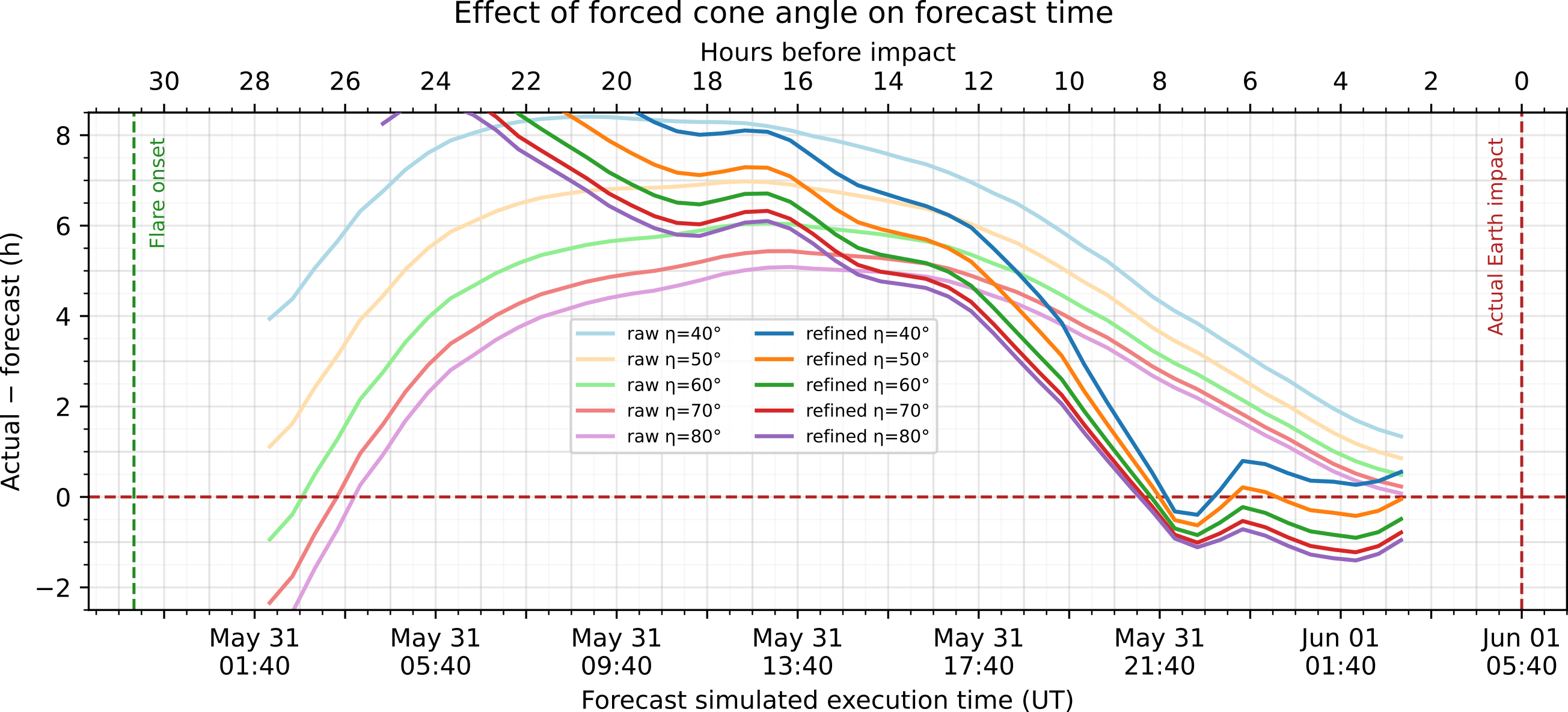}
    \caption{Ensemble forecasts of the same 2025 May 31 CME, with $t_0$ held fixed and
    $\eta$ varying between forecasts, show that the refined forecasts are moderately 
    insensitive to the
    forced $\eta$ (half-cone) angle.  The arbitrary selection of $60^\circ$ for the 
    primary forecast series in Figure \ref{fig:forecasts}
    does affect the arrival-time prediction, but only weakly.  The entire ensemble of refined
    predictions remains within $\pm 1.5$ hours of the actual arrival time, beginning 8
    hours before impact.}
    \label{fig:eta_scan}
\end{figure}

Figure \ref{fig:eta_scan} is an overlay of five instances of the process plotted in the top
panel of Figure \ref{fig:forecasts}, with values of $\eta$ ranging from 40$^\circ$ to 
80$^\circ$. The initial phases of the fit 
yield different values, because of the 
small-angle degeneracy between $\eta$ and $v_{CME}$.  As the ``lever arm'' shrinks, the 
entire ensemble of forecasts converge (all roughly 8 hours prior to impact) to similar, 
but slightly different, values.  The convergence point is a clear inflection, and could
be identified in any model in the ensemble from the rapid change in the derivative of
consecutive prediction times with respect to forecast time.
The best of the ensemble (determined post-hoc) is $\eta=70^\circ$, which, starting at 8 hours before impact, 
remains within 15 minutes of the actual impact time until the CME leaves the field of view.  This is a good reminder that the arbitrary choice of $\eta=60^\circ$ was 
not optimized for the event.

\begin{figure}
    \centering
    \includegraphics[width=0.9\linewidth]{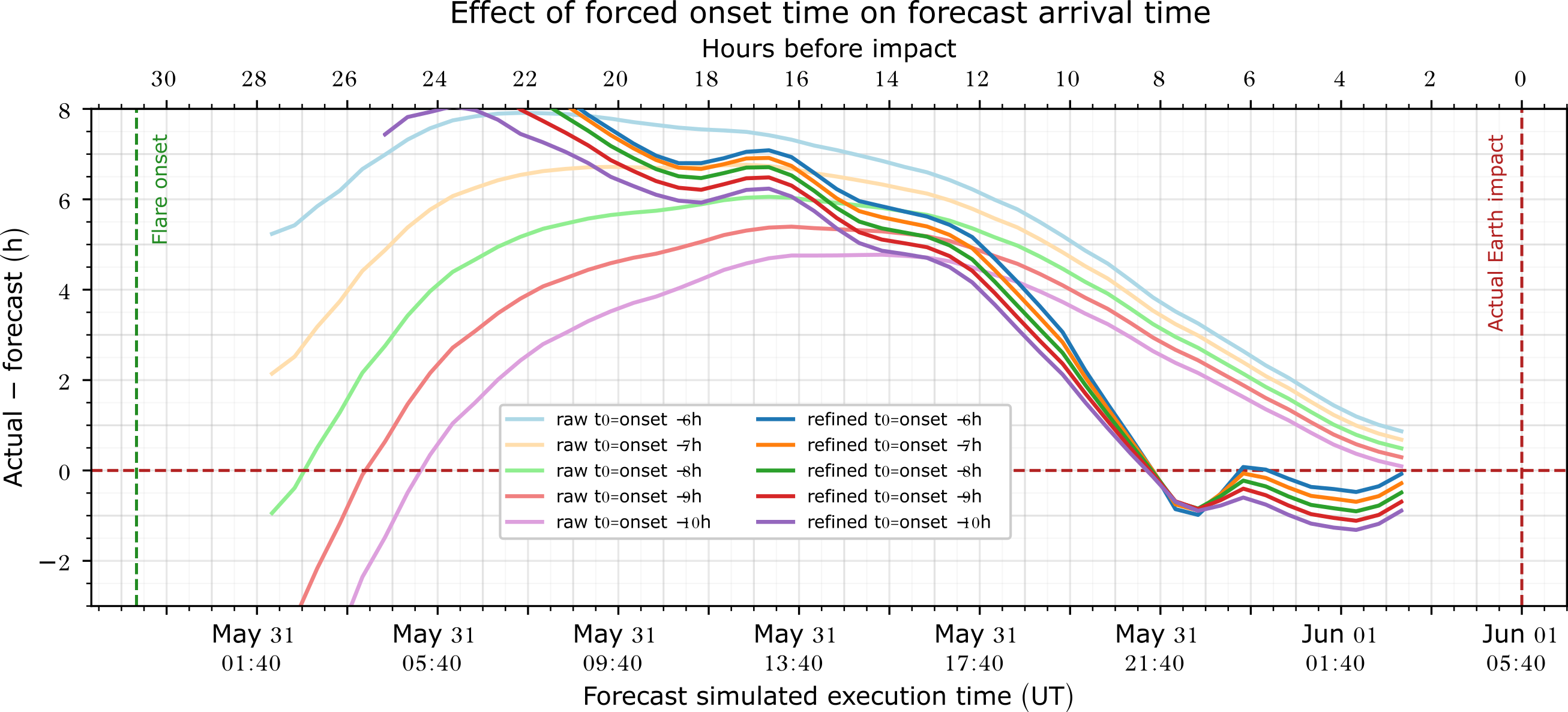}
    \caption{Ensemble forecasts of the same 2025 May 31 CME, with $\eta$ held fixed and
    $t_0$ varying between forecasts, show that the refined forecasts are moderately 
    insensitive to the forced $t_0$ (pre-eruption to account for braking) time.  The
    arbitrary selection of 8 hours pre-flare for $t_0$ in the primary forecast series in 
    Figure \ref{fig:forecasts} does affect the arrival-time prediction, but only weakly.  The
    entire ensemble of refined predictions remains within $\pm$ 1.5 hours of the actual arrival time, beginning 8 hours before impact. }
    \label{fig:t0_scan}
\end{figure}

Figure \ref{fig:t0_scan} is an overlay of five instances of the process plotted in the top 
panel of Figure \ref{fig:forecasts}, with values of $t_0$ ranging from 10 hours before the flare, to 6 hours before the flare.  The initial phases of the fit yield wildly different
values, because the constant-velocity extrapolation greatly overestimates average propagation
speed when informed only by early tracking data.  The fits converge to nearly constant values, 
with different arrival times that are modified by the additional allowed propagation time when 
the model is informed only by late tracking data.  The offset launch time offsets the 
convergence time by an amount proportional to the onset shift, scaled to the remainder of the
propagation time.  

In sum: correcting $\eta$ to lower values than the initial $90^\circ$ to which free fits 
were seen to converge makes the final predictions \textit{earlier}, while correcting $t_0$ to 
earlier values makes the final predictions \textit{later}.  The two effects approximately 
canceled in the initial forecast, but in either case are small compared to the late-phase
gains from continuous tracking.  While the choice of particular $\eta$ and $t_0$ offsets
certainly affects the predictions of the toy model, that effect is small compared to the
fundamental gain from continuous tracking of the CME front.

\section{Discussion and Conclusions} \label{sec:conclusion}

We have demonstrated a natural extension of simple geometric front-shape measurement to 
the large field of 
view and few-minute imaging cadence afforded by the PUNCH mission.  
Wide-field CME tracking was first performed with the Helios mission 
\cite{jackson_leinert_1985}, and later refined with Coriolis/SMEI 
and STEREO/SECCHI; the novelty of this
preliminary data set from PUNCH is in its rapid cadence and low noise (even in 
the preliminary state) compared to data
from SMEI, and full-circle aspect compared to SECCHI. These characteristics 
allowed us to track the 
CME of 2025-May-31 continuously across most of its propagation from Sun to Earth.

We used simple rotoscoping (point-and-click estimation) of the leading edge.  
We found that the geometric front of the CME evolves
complexly, corroborating prior inferences that at least some CMEs are causally
disconnected internally, and therefore develop structure as they evolve in the 
inner heliosphere. This complexity is important to understanding the physics of CME 
propagation, and forms a source of error for simplified models that treat the system as
an analytic, smooth, or simple and internally coherent structure.  

For the remainder of this
analysis, we studied the utility of tracking to retrospectively forecast the arrival 
time of the CME.  
We deliberately ignored the observed internal structure, to isolate the benefits 
of continuous tracking from the presumed benefits of more 
detailed structural
insight.  We fitted the complicated halo CME leading edge with the simplest of simple models:  a 
constant-velocity spherical ice-cream cone shape with fixed (prescribed) cone angle 
$\eta$ and start time $t_0$, leaving only three free parameters (the fixed front-face speed
$v_{nose}$, the off-Earth-Sun angle $\theta$, and the sky-plane directional argument 
$\alpha_0$).  This ice-cream cone model was sufficient to roughly characterize the CME and achieve 
a visual match between the complex leading-edge structure in the PUNCH data, and the overall
visual form of the halo.  

We demonstrated using this very simple model to ``forecast'' CME arrival 
time, via analytic constant-velocity extrapolation/propagation from data subsets prior 
to prescribed forecast intervals. The technique of using tracking across the inner 
solar system is 
sufficiently powerful that even this unsophisticated analytic shape model 
converged on a clean forecast of the CME arrival time.  Using a straightforward refinement 
(linearization of the forecast-vs-time response of the ice-cream cone extrapolations, and
extrapolation to refine the current prediction based on prior behavior of the model itself) 
yielded forecast accuracy of a fraction of an hour, from 8 hours prior to impact onward.  This
model compared favorably with values from the competitive ``CME Scoreboard'' operated by the 
Community
Coordinated Modeling Center to validate state-of-the-art forecasting 
models \cite{riley_etal2018}.  
Across the
11 entries for this event, the RMS forecast error was 5 hours \cite{ccmc_cme_scoreboard_2025};
the ``PUNCH and ludicrous ice-cream cone'' model described here performed an order of magnitude
better than this benchmark.  That result is both promising and (as a single 
retroforecast result) anecdotal. It primarily indicates the importance of follow-up work to understand the limits of the method.

The fact that the analytic forecast fits converged so readily to the actual measured
arrival time, even with initial data
collected during spacecraft commissioning and a toy semi-analytic geometric model, 
highlights the importance of 
rapid, wide-field imaging to improve understanding of CME geometry and propagation. Including
the photometric and polarimetric improvements from later data releases is anticipated to 
yield direct CME localization in 3-D (via polarimetry) and structural evolution data (via 
high-SNR imagery). We have demonstrated here that even crude and direct morphological 
tracking
with moderate spatial resolution and few-minute cadence, over the wide PUNCH field of view,
is sufficient to capture and forecast the propagation of a halo CME from Sun to Earth.

Because the simplified model had two important parameters ($\eta$ and $t_0$) that were 
picked arbitrarily, we investigated the effect of picking different values on the final
forecast quality.  We found that the choice of $\eta$ and $t_0$ does affect error in 
arrival time predictions, but only weakly. While more detailed study is needed to optimize the choice, all the choices we considered offered better prediction skill than the more 
conventional extrapolation
methods on the CCMC scoreboard for this event.

The refinement method we used, calculating a linearly extrapolated 
``forecast-of-forecasts'', not only improves the forecasted outcome, it is stable against
a broad range of errors in the geometric forecast model itself.  Wide-field imaging affords 
tracking events nearly all the way to impact with Earth, and later observations yield
successively more (and more precise) information about that impact, both because
perspective effects cause a steepening of the elongation-vs-time curve for the front 
(as in Figure \ref{fig:model-scans}), and because the remaining propagation time grows
shorter with each observation. Provided that the analytic shape model can in principle
capture the $\varepsilon=90^\circ$ moment of impact, and that the $\varepsilon$ vs. $\ell_{CP}$ curve varies smoothly, the forecast time-of-impact must also converge 
smoothly to the 
actual arrival time, as that arrival time approaches. Linear extrapolation of the 
forecast time uses the 
structure of the intrinsic error function of the geometric model itself and the 
necessary convergence of the forecast time to the present.  The sharp ``knee'' we
observed in Figure \ref{fig:forecasts} represents the beginning of the regime in which
the convergence of raw forecast error to zero may be approximated with a straight
line.

Notably, the forecast-of-forecasts approach is not dependent on the arrival cadence 
of the data, only on the acquisition cadence.  In a genuine forecast situation during
nominal
science operations, the PUNCH/WFIs would have made between 1 and 2 complete sets of 
ground passes (between 3 and 6 ground passes total) over the course of the event itself, 
affording only up to about six chances to refine a forecast during the event propagation
interval.  But the detrending refinement method, using 30-minute forecasts would 
still be feasible, by executing multiple forecasts using subsets of the downlinked data
in each tranche.

The power of wide-field tracking and modeling to greatly improve forecast
arrival times has been demonstrated before, by \citeA{webb_etal2009b}.  They 
used refined tracks of three CMEs seen by the SMEI spacecraft, and demonstrated that
the best of their retroforecast times was approximately 30 minutes from the measured
arrival time of the CME at Earth.  The novelty of the current analysis is twofold: (1) 
PUNCH data are more frequent, showing the effect of forecast updates at a twice-hourly cadence, rather than 
(as with \citeauthor{webb_etal2009b}) once every three hours; and (2) we
demonstrate a clearly identifiable feature in the forecast trend plot (the inflection 
point in the refined arrival-time forecasts), which identifies convergence of the model 
in simulated real time and was not available with the lower-cadence SMEI data.

The enticingly strong outcome of the work is a proof of concept,
demonstrating via a single case that high quality, high cadence wide-field data can, in principle, generate greatly improved CME arrival time
forecasts with continuous 
refinement, even with simple data analysis and crude forecast modeling.  Ensemble testing, ideally 
culminating in live forecasts using the QuickPUNCH data stream \cite{seaton_etal_2025} on the CCMC 
scoreboard or similar public forum, is necessary to understand the limits of that result.  Nevertheless,
the method is promising.  There are many directions to improve: higher data quality using 
later, less preliminary versions of the mission's 
data products; less ludicrously simplified modeling (possibly informed by more extensive analysis of solar surface, coronal, and near-Sun space data); automated front capture; 
inference from CME front structure or interior structure; and polarimetric analysis to 
locate the front directly in three dimensions \cite<e.g.,>{deforest_etal_2013, gibson_etal_2025}
are all elements that may be expected to 
improve the forecast quality and lead time over this proof-of-concept study.  We 
have demonstrated here that full-circle tracking across the solar system is sufficiently powerful to overcome 
the limitations of even crude analysis and extrapolation methods; and that sufficiently
high cadence data can yield a precise arrival-time forecast, hours ahead of impact, 
with an immediately-available indicator of validity.


\section*{Open Research}

We used data from the following sources.

\begin{description}
    \item[PUNCH]{The primary data set for this analysis was a PUNCH sample movie rendered from the early (commissioning-phase) PUNCH data.  All PUNCH data products are public.  The frames are available at the Zenodo repository \cite{deforest_etal_2026c}; the movie
    is attached to the Supplementary Information for this article; and the original level-0 data frames are archived at NASA's Solar Data Analysis center (https://umbra.nascom.nasa.gov/punch).}
    
    \item[SOHO]{This analysis used data from the SOHO/LASCO instrument.  LASCO data are archived at the Naval Research Laboratory and may be retrieved via (https://lasco-www.nrl.navy.mil) or from the SOHO archive (https://soho.nascom.nasa.gov/data/archive).}

    \item[GOES]{This analysis used data from the GOES X-ray irradiance sensor.  Figure \ref{fig:goes} was rendered using the SWx TREC Space Weather data
    portal \cite{SWxTREC2019}.}

    \item[OMNI]{This analysis used 5-minute OMNI data to describe the near-Earth solar wind environment.  OMNI data may be obtained from the NASA/GSFC Space Physics Data Facility's OMNIWeb service (https://omniweb.gsfc.nasa.gov/).}

    \item[ACE]{This analysis used data from the ACE mission.  ACE data may be downloaded from NASA's Coordinated Data Analysis Web (https://cdaweb.gsfc.nasa.gov).}
\end{description}

\section*{Conflicts of Interest}

The authors declare that they have no conflicts of interest.

\acknowledgments
We wish to acknowledge the Boulder Solar Alliance Research Experience for Undergraduates program,
who enabled KAR to work on this project in the summer of 2025. SEG acknowledges the National Center for Atmospheric Research, a major facility sponsored by the National Science Foundation under Cooperative Agreement No. 1852977.  The authors also wish
to thank R. Colaninno, T. Amerstorfer, and the two anonymous referees for review and comments that improved the work.
PUNCH is a heliophysics mission to study the corona, solar wind, and space weather as an integrated system,
and is part of NASA's Explorers program (Contract 80GSFC14C0014).

\bibliography{ms}

\end{document}